\documentclass[lineno]{jfm}

\usepackage{graphicx}
\usepackage{newtxtext}
\usepackage{newtxmath}
\usepackage{natbib}
\usepackage{hyperref}
\hypersetup{
    colorlinks = true,
    urlcolor   = blue,
    citecolor  = black,
}
\newcommand{\RomanNumeralCaps}[1]
\linenumbers

\title{From Sheared Annular Centrifugal Rayleigh-Bénard Convection to Radially Heated Taylor-Couette Flow: Exploring the Impact of Buoyancy and Shear on Heat Transfer and Flow Structure}

\author{Jun Zhong\aff{1},
  Dongpu Wang\aff{1}
 \and Chao Sun\aff{1,2}
 \corresp{\email{chaosun@tsinghua.edu.cn}}}

\affiliation{\aff{1} Center for Combustion Energy, Key Laboratory for Thermal Science and Power Engineering of Ministry of Education, and Department of Energy and Power Engineering, Tsinghua University, 100084 Beijing, China
\aff{2} Department of Engineering Mechanics, School of Aerospace Engineering, Tsinghua University, 100084 Beijing, China}

\begin{document}
\maketitle

\begin{abstract}
We investigate the coupling effect of buoyancy and shear based on an annular centrifugal Rayleigh-B\'enard convection (ACRBC) system in which two cylinders rotate with an angular velocity difference. Direct numerical simulations are performed in a Rayleigh number range $10^6\le Ra\le 10^8$, at fixed Prandtl number $Pr=4.3$, inversed Rossby number $Ro^{-1}=20$ and radius ratio $\eta=0.5$. The shear, represented by the non-dimensional rotational speed difference $\Omega$, varies from $0$ to $10$, corresponding to an ACRBC without shear and a radially heated Taylor-Couette flow with only the inner cylinder rotating, respectively. A stable regime is found in the middle part of the interval of $\Omega$, and divides the whole parameter space into three regimes: buoyancy-dominated regime, stable regime, and shear-dominated regime. Clear boundaries between the regimes are given by linear stability analysis, meaning the marginal state of the flow. In our parameter space, we find that the marginal state is sensitive to $Ra$ at a low Rayleigh number while mainly depending on $\Omega$ when $Ra\ge 10^6$. In the buoyancy-dominated regime, the flow is a quasi-two-dimensional flow on the $r\varphi$ plane; as shear increases, both the growth rate of instability and the heat transfer is depressed. Firstly, the number of convection roll pairs decreases; then the convection roll disappears with heat mainly transferred by strong plumes; finally, plumes disappear as well. In the shear-dominated regime, the flow is mainly on the $rz$ plane, first driven by buoyancy and then taken over by shear quickly as $\Omega$ increases. The shear is so strong that the temperature acts as a passive scalar, and the heat transfer is greatly enhanced. The study shows shear can stabilize buoyancy-driven convection, makes a detailed analysis of the flow characteristics in different regimes, and reveals the complex coupling mechanism of shear and buoyancy, which may have implications for fundamental studies and industrial designs.
\end{abstract}



\section{Introduction}

Turbulent convection, one of the most complicated fluid motions, is ubiquitous in nature and industrial processes. As two essential elements of fluid dynamics, buoyancy, and shear play important roles in many kinds of turbulent flows. Under gravity or other body force fields, buoyancy is generated by the inhomogeneous density distribution of fluid. When the buoyancy force is large, it can induce instability and drive the convection. Rayleigh-Bénard (RB) convection is one typical paradigm of buoyancy-driven convection and has been studied extensively in scientific research \citep{ahlers_heat_2009,lohse_small-scale_2010,chilla_new_2012,xia_current_2013}. In the RB cell, the fluid is heated from below and cooled from above under gravity. Due to the thermal expansion of the fluid, buoyancy appears and drives the convection. Some manifold and involute flow structures are formed in the cell \citep{niemela_wind_2001,xi_laminar_2004,sun_three-dimensional_2005,wang_heat_2021}. Moreover, apart from the classical RB model with rectangular cells, annular \citep{pitz_onset_2017,kang_numerical_2019,jiang_supergravitational_2020,rouhi_coriolis_2021,wang_statistics_2023} and spherical \citep{gastine_turbulent_2015} RB cells also attract a lot of interest. The uniform gravity is substituted by the centrifugal force or gravity which varies in the direction of the temperature gradient.  Different from buoyancy, shear contributes to the flow mainly by the motion of the system boundary. Taylor-Couette (TC) flow, the flow impelled by two concentric cylinders rotating independently, is a widely used canonical model to study the effect of shear. Secondary flows are caused by the centrifugal instability, and then many complicated flow structures including turbulent Taylor vortex flow and wavelets are generated at high Reynolds numbers \citep{bayly_threedimensional_1988,esser_analytic_1996,brauckmann_intermittent_2013, ostilla-monico_optimal_2014,grossmann_highreynolds_2016}.

Transport efficiency is an important quantity in both RB convection and TC flow, as the physical quantities being transferred are temperature and angular velocity, respectively \citep{eckhardt_torque_2007}. In non-dimensional forms, there exist scaling laws in RB convection, between the dimensionless heat flux (measured by the Nusselt number $Nu_h$) and the dimensionless buoyancy-driven strength (measured by the Rayleigh number $Ra$) \citep{castaing_scaling_1989,shraiman_heat_1990,grossmann_scaling_2000}; similar scaling laws occur in TC flow as well, between the dimensionless angular velocity current (measured by $Nu_\omega$) and the dimensionless shear (measured by the Taylor number $Ta$) \citep{van_gils_torque_2011,merbold_torque_2013}. Bradshaw observed The high similarity between the RB convection and TC flow \citep{bradshaw_analogy_1969}; further, an exact analogy between the two flows is raised, extending the Grossmann \& Lohse's scaling theory on RB convection well to TC flow \citep{eckhardt_torque_2007,busse_twins_2012}. This analogy reveals that the transport phenomena in the two systems have similar inner physics.

The coupling of buoyancy and shear is widespread in atmospheric motion and oceanic flow \citep{Dearorff_Numerical_1972, Khanna_three_1998}. For example, the surface mesoscale eddies in the ocean are mainly generated by baroclinic instabilities, which are relevant to the pole-equator temperature gradient and a vertical shear \citep{Hopfinger_Vortices_1993, Pierrehumbert_bar_1995,Egbers2014, Feng_Season_2022}. By far, there are many attempts to combine shear and buoyancy in one system and study their coupling effect. When a radial temperature difference is applied to a TC system, it contributes to the instability \citep{yoshikawa_instability_2013,meyer_effect_2015,kang_thermal_2015} and the momentum and heat transfer \citep{kang_radial_2017,leng_mutual_2022}. When the gravity or centrifugal force is considered as the body force inducing buoyancy, the results are different. Leng et al. \citep{leng_flow_2021,leng_mutual_2022} apply radial or axial temperature difference on TC flow with only the inner cylinder rotating, and find that with fixed temperature difference and increasing rotational speed, the heat transfer is first depressed by shear and then enhanced due to the development of turbulent TC flow. Interestingly, similar phenomena occur when shear is applied on a rectangular RB cell \citep{goluskin_convectively_2014,blass_flow_2020,blass_effect_2021}. With increasing plane Couette shearing, the flow is first dominated by buoyancy and then dominated by shear. The heat transport shows a similar trend in the two systems, that heat transport is depressed under weak shearing until the shear is strong enough to mix the system better than thermal plumes. The transition of dominated regimes depends on the Richardson number $Ri$, defined as the ratio between buoyancy and shear driving. 

To explore the effect of shear on heat transfer in a wide parameter regime, we adopt the annular centrifugal Rayleigh-Bénard convection (ACRBC). The centrifugal force generated by rotating inner and outer cylinders offers buoyancy and the Earth's gravity is neglected, similar to the rotating machines with high rotational speed. As the buoyancy is perpendicular to the cylinder surface, the aspect ratio (here, the aspect ratio is defined as the ratio of circumference to the gap length) in ACRBC is larger than that in RB convection. Recent studies show that the ACRBC owns similar scaling laws with classical RB convection, and is an efficient way to reach the ultimate regime \citep{jiang_supergravitational_2020,jiang_experimental_2022}. The angular velocity difference between the inner and outer cylinders offers shear for the system, therefore the system can also be regarded as a Taylor-Couette system with radial temperature difference. With no shear, the inner and outer cylinders co-rotating at the same angular speed; with strong shear, one of the cylinders can stop rotating or become counter-rotating with another. \cite{meyer_effect_2015}  discusses the instability of a similar system with a non-rotating outer cylinder but mainly focuses on the instability at a low Taylor number. In our system, we work on a wide range of the Rayleigh number and the Taylor number, concentrating on the instability and scalar transport, aiming to bring a complete understanding of the coupling effect of buoyancy and shear in an annular centrifugal system.

The rest of the paper is organized as follows: the establishment of the numerical model and the numerical methods including linear stability analysis (LSA) and direct numerical simulation (DNS) are introduced in section \ref{sec2}, and the main results are discussed in section \ref{sec3}. Finally, conclusions are presented in section \ref{sec4}.

\section{Numerical model}\label{sec2}

\begin{figure}
    \centering
    \includegraphics[width=0.7\linewidth]{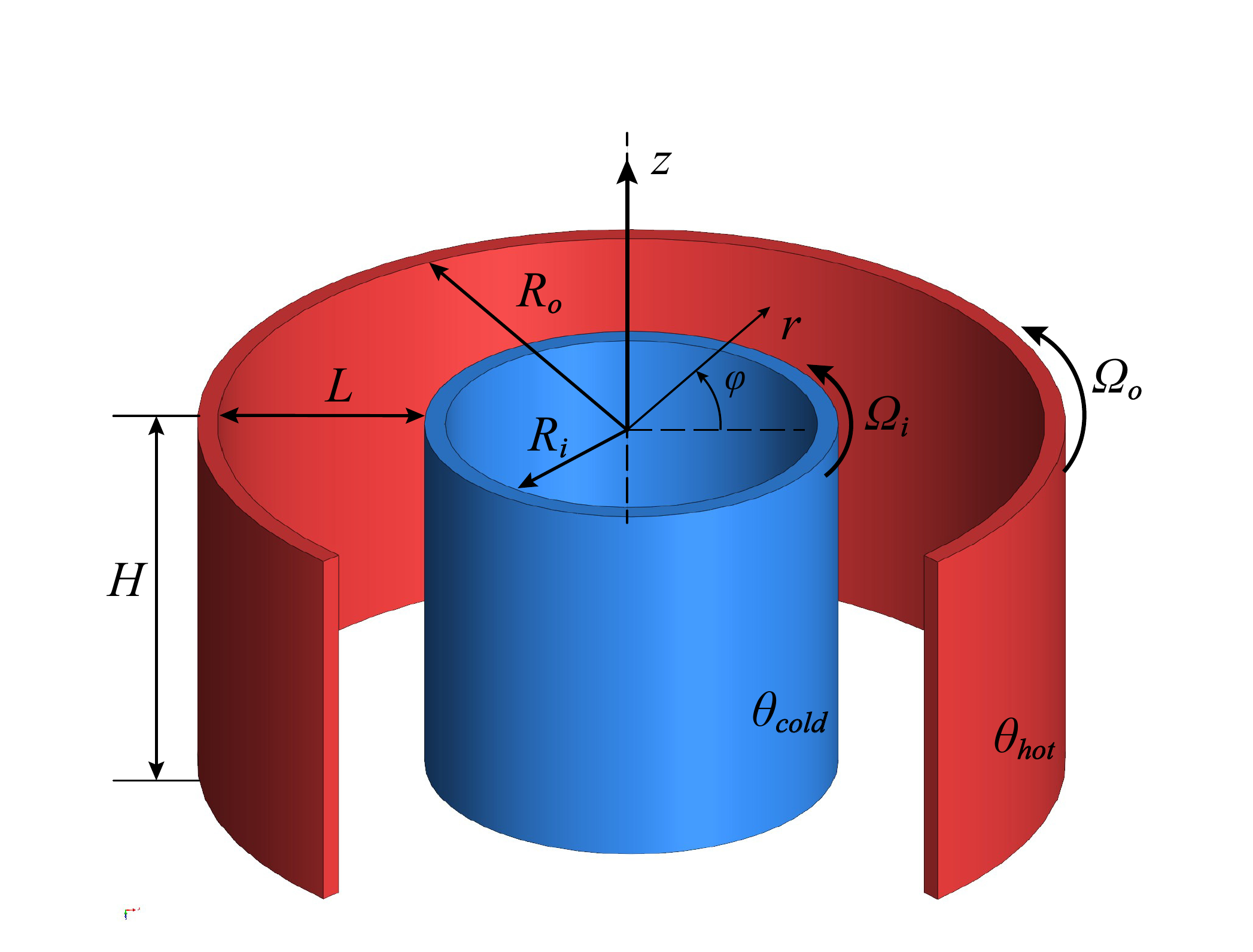}
    \captionsetup{justification=raggedright}
    \caption{ Schematic diagram of the flow configuration in the stationary reference frame. $R_o$,$R_i$, and $L$ are the inner radius of the outer cylinder, the outer radius of the inner cylinder, and the gap width between the two cylinders, respectively. $H$ is the height of the cylindrical annulus in the computational domain of DNS. The outer cylinder rotates at the angular velocity $\Omega_o$ and the inner cylinder rotates at the angular velocity $\Omega_i$. $\theta_{hot}$ and $\theta_{cold}$ are the temperature of the outer and inner walls.}
	\label{fig:Schema}
\end{figure}

A three-dimensional annular centrifugal RB cell bounded by two independent-rotating concentric cylinders is considered, as shown in figure \ref{fig:Schema}. The inner cylinder with radius $R_i$ rotates about $z$ axis at angular velocity $\Omega_i$ and the outer cylinder with radius $R_o$ rotates at the angular velocity $\Omega_o$. $L=R_o-R_i$ is the gap width between the two cylinders. The temperature difference between the hot outer cylinder and the cold inner cylinder is $\Delta=\theta_{hot}-\theta_{cold}$. For the boundary, no-slip and isothermal conditions are applied at two cylinder surfaces; periodic boundary conditions are imposed on the velocity and temperature in the axial direction, and we take a section of height $H$ as the computational domain in DNS.

\subsection{Governing equations of the flow}

The system is at a rotating frame with angular velocity $\Omega_c$, and the buoyancy is induced by the centrifugal force $(\Omega_cr+u_{\varphi})^2/r$. The motion of the flow is governed by the non-dimensional Oberbeck–Boussinesq equations under cylindrical coordinate system $(r,\varphi,z)$ \citep{jiang_supergravitational_2020,wang_effects_2022}:

\begin{equation}\label{OBequ}
    \begin{aligned}
    \boldsymbol{\nabla}\cdot\boldsymbol{u}&=0,\\
    \frac{\partial\boldsymbol{u}}{\partial t}+\boldsymbol{u}\cdot\boldsymbol{\nabla u}=-\boldsymbol{\nabla} p-Ro^{-1}\boldsymbol{e_z}\times\boldsymbol{u}&+\sqrt{\frac{Pr}{Ra}}\nabla^2\boldsymbol{u}-\theta\frac{2(1-\eta)}{1+\eta}(1+\frac{2u_\varphi}{Ro^{-1}r})^2\boldsymbol{r},\\
    \frac{\partial \theta}{\partial t}+\boldsymbol{u}\cdot\boldsymbol{\nabla} \theta&=\sqrt{\frac{1}{Ra\cdot Pr}}\nabla^2 \theta,\\
    \end{aligned}
\end{equation}
where $\boldsymbol{u}=(u_r,u_\varphi,u_z)$ is the velocity vector, $\theta\in[0,1]$ is the temperature, $p$ is the pressure, $\boldsymbol{e_z}$ is the unit vector along the axial direction and $\eta=R_i/R_o$ is the radius ratio. Scaled quantities, including $L=R_o-R_i$ for length, $\Delta$ for temperature, $U=\sqrt{\alpha\Delta \Omega_c^2\frac{(R_i+R_o)}{2}L}$ for velocity, and $L/U$ for time are used to non-dimensionalize the governing equation, where $\alpha$ is the coefficient of thermal expansion of the fluid. $\Omega_c$ is the rotational speed of the rotating reference frame, and we suppose it to represent the strength of centrifugal buoyancy, as reflected in the expression of free-fall velocity $U$. In our system, considering both physical meaning and simplicity, we set $\Omega_c=(\Omega_i+\Omega_o)/2$. It is a good estimate for free-fall velocity when shearing is relatively small, which means $\Omega_i-\Omega_o\ll\Omega_c$; when shearing is strong enough and dominates the flow, buoyancy contributes little and can be neglected, which will be discussed later in section \ref{sec3}. Therefore, the selected $\Omega_c$ is reasonable in most of the parameter space. Under that rotating frame, the inner cylinder rotates at non-dimensional angular velocity $\Omega=(\Omega_i-\Omega_c)L/U$ and the outer cylinder rotates at $-\Omega$.

Three non-dimensional parameters are generated through non-dimensionalization, they are the Rayleigh number (measuring the buoyancy-driving strength) $Ra$, the inverse Rossby number (measuring Coriolis effects) $Ro^{-1}$ and the Prandtl number (fluid property) $Pr$, expressed as:

\begin{equation}\label{ParEqu}
    \begin{aligned}
        Ra=\frac{(UL)^2}{\nu\kappa}&=\frac{\alpha\Delta\Omega_c^2\frac{(R_i+R_o)}{2}L^3}{\nu\kappa},\\
        Ro^{-1}=\frac{2\Omega_cL}{U}&=2(\frac{\alpha\Delta(R_i+R_o)}{2L})^{-1/2},\\
        Pr&=\frac{\nu}{\kappa},
    \end{aligned}
\end{equation}
where $\nu$ is the kinematic viscosity and $\kappa$ is the thermal diffusivity of the fluid. Therefore, there are five independent control parameters in the system: the Rayleigh number $Ra$, the inverse Rossby number $Ro^{-1}$, the Prandtl number $Pr$, the rotational speed difference of two cylinders $\Omega=(\Omega_i-\Omega_c)L/U$ and the radius ratio $\eta$. Moreover, since the system can also be regarded as a TC system with radial buoyancy, the Taylor number $Ta$ can be defined as:
\begin{equation}\label{TaEqu}
    Ta=\frac{(1+\eta)^4}{64\eta^2}\frac{(R_o-R_i)^2(R_i+R_o)^2(\Omega_i-\Omega_o)^2}{\nu^2}=\frac{(1+\eta)^6}{16\eta^2(1-\eta)^2}\frac{\Omega^2Ra}{Pr}.
\end{equation}
Therefore, $Ta$ is not an independent control parameter in our system, proportional to the $Ra$ and $\Omega^2$. The Richardson number $Ri$, measuring the ratio between buoyancy and shear driving, can also be calculated by the ratio of characteristic buoyancy-velocity and the characteristic shearing-velocity:
\begin{equation}\label{RiEqu}
    Ri=\frac{U^2}{(\Omega_i-\Omega_c)^2r_i^2}=(\frac{1-\eta}{\eta})^2\Omega^{-2}.
\end{equation}
As we can see, the non-dimensional rotational speed difference $\Omega$ can represent the strength of shear relative to buoyancy.

In addition, two key response parameters are the Nusselt numbers measuring the efficiency of heat transport and momentum transport, given by the ratio of radius-independent currents to the currents in laminar and nonvortical cases, respectively \citep{eckhardt_torque_2007,wang_effects_2022}, as $Nu_h=J^\theta/J^\theta_{lam}$, $Nu_\omega=J^\omega/J^\omega_{lam}$.  More about the Nusselt numbers and how they respond to $Ra$ and $\Omega$ will be discussed in section \ref{sec3}.


\subsection{Linear stability analysis}

LSA is a good approach to studying the instability of the flow. At the stable state, the flow is laminar and nonvortical TC flow and the heat is transferred by pure conduction. The azimuthal velocity $V$ and temperature $\Theta$ depend only on $r$, and the radial and axial velocities are zero. The functions of $V$ and $\Theta$ are given by \citep{ali_stability_1990}
\begin{equation}\label{LamEqu}
    \begin{aligned}          
    V(r)=Ar+\frac{B}{r},  A&=-\frac{1+\eta^2}{1-\eta^2}\Omega,  B=\frac{2r_i^2}{1-\eta^2}\Omega,\\
    \Theta(r)&=\frac{ln(r/r_i)}{ln(r_o/r_i)},\\
    \end{aligned}
\end{equation}
where $r_i=\eta/(1-\eta)$ and $r_o=1/(1-\eta)$ are the non-dimensional radii of the inner and outer cylinders, respectively. 

To perform LSA on this problem, we superimpose infinitesimal perturbations $(u'_r,u'_\varphi,u'_z,p',\theta')$ on the base flow state. After substituting the perturbation fields into equations (\ref{OBequ}), doing linearization, and expanding the perturbations into normal modes \citep{meyer_effect_2015,kang_radial_2017},
\begin{equation}\label{equ：decomp}
    (u'_r,u'_\varphi,u'_z,p',\theta')=(\hat{u_r}(r),\hat{u_\varphi}(r),\hat{u_z}(r),\hat{p}(r),\hat\theta(r))exp(st+i(n\varphi+kz)),
\end{equation}
we obtain the resulting ordinary differential equations for $r-$ dependent normal mode quantities:
\begin{equation}\label{LSAequ}
    \begin{aligned}
        (D+r^{-1})\hat{u_r}+&inr^{-1}\hat{u_\varphi}+ik\hat{u_z}=0,\\
        (s+\frac{inV}{r})\hat{u_r}-\frac{2V}{r}\hat{u_\varphi}=&-D\hat{p}+Ro^{-1}\hat{u_\varphi}+\sqrt{\frac{Pr}{Ra}}(\nabla^2\hat{u_r}-\frac{\hat{u_r}}{r^2}-\frac{2in\hat{u_\varphi}}{r^2})\\
        &-\frac{2(1-\eta)}{(1+\eta)}r[(1+\frac{2V}{Ro^{-1}r})^2\hat{\theta}+\frac{4\Theta}{Ro^{-1}r}(1+\frac{2V}{Ro^{-1}r})\hat{u_\varphi}],\\
        (s+\frac{inV}{r})\hat{u_\varphi}+(DV+\frac{V}{r})\hat{u_r}=&-\frac{in}{r}\hat{p}-Ro^{-1}\hat{u_r}+\sqrt{\frac{Pr}{Ra}}(\nabla^2\hat{u_\varphi}-\frac{\hat{u_\varphi}}{r^2}+\frac{2in\hat{u_r}}{r^2}),\\
        (s+\frac{inV}{r})\hat{u_z}=&-ik\hat{p}+\sqrt{\frac{Pr}{Ra}}\nabla^2\hat{u_z},\\
        (s+\frac{inV}{r})\hat{\theta}+(D\Theta)\hat{u_r}=&\frac{1}{\sqrt{RaPr}}\nabla^2\hat{\theta},
    \end{aligned}
\end{equation}
where operators $D=d/dr$ and $\nabla^2=D^2+D/r-n^2/r^2-k^2$ are introduced for simplification. $s$ is the temporal growth rate of perturbation, $n$ is the azimuthal mode number and $k$ is the axial wavenumber. Due to the infinite axial length and $2\pi$ period in the azimuthal direction, $k$ must be real and $n$ must be an integer. The boundary conditions of the perturbations are homogeneous:
\begin{equation}\label{LSABL}
    \begin{aligned}   
    r=r_i: \hat{u_r}&=\hat{u_\varphi}=\hat{u_z}=\hat{p}=\hat\theta=0,\\
    r=r_o: \hat{u_r}&=\hat{u_\varphi}=\hat{u_z}=\hat{p}=\hat\theta=0.
    \end{aligned}
\end{equation}

Then, the instability problem has been transformed into an eigenvalue problem, described by equations (\ref{LSAequ})-(\ref{LSABL}). This eigenvalue problem is solved by Chebyshev spectral collocation method. The equations (\ref{LSAequ}) are discretized on Chebyshev-Gauss-Lobatto collocation points with Chebyshev differentiation matrices. In our work, the number of collocation points range from $200$ to $400$ for good convergence. Then the growth rate of perturbations $s$ becomes the eigenvalue of the generalized eigenvalue problems in the matrix form, and the corresponding perturbation normal modes are the eigenvectors. For certain parameters $(Ra, Pr, Ro^{-1}, \Omega, \eta)$, the flow is stable if for all $(n, k)$, the growth rate $\sigma=real(s)$ is always negative, which means all perturbation modes decay with time.
   
\subsection{Direct numerical simulation}

Direct numerical simulations (DNS) are performed using an energy-conserving second-order finite-difference code based on a Chebyshev-clustered staggered grid. The time-stepping of the explicit terms is based on a fractional-step third-order Runge–Kutta scheme, and the implicit terms are based on a Crank–Nicolson scheme with a pressure correction step set following. For more details on the numerical schemes of the governing equations, we refer the reader to this literature \citep{verzicco_finite-difference_1996,van_der_poel_pencil_2015,zhu_afid-gpu_2018}. 

Adequate resolutions are ensured for all simulations and we have performed posterior checks of spatial and temporal resolutions to guarantee the resolution of all relevant scales. The ratios of maximum grid spacing $\Delta_g$ to the Kolmogorov scale estimated by the global criterion $\eta_K=(\nu^3/\varepsilon)^{1/4}$, where $\varepsilon$ is the mean viscous dissipation rate via exact relation, and the Batchelor scale $\eta_B=\eta_KPr^{-1/2}$ \citep{silano_numerical_2010} are checked, as shown in Appendix \ref{Tab1}. Furthermore, the clipped Chebychev-type clustering grids adopted in the radial direction ensure the spatial resolution within boundary layers (BLs), as at least $10$ grid points inside the thermal boundary layers. As for temporal resolution, we use the Courant-Friedrichs-Lewy (CFL) conditions and set $CFL\le 0.7$ to guarantee the computational stability \citep{ostilla_optimal_2013,van_der_poel_pencil_2015,zhang_statistics_2017}. The simulations are run over enough time after the system has reached the statistically stationary state to obtain good statistical convergence. The relative difference of $Nu_h$ based on the first and second halves of the simulations is generally less than $1\%$. And for $Nu_\omega$, as its absolute value is small and close to $1$ in many weak shear cases, the relative difference is controlled generally less than $4\%$ under weak shear, and less than $2\%$ under strong shear. All those details are provided in Appendix \ref{Tab1}.

\subsection{Other numerical details}

In the present study, we aim at the coupling effect of shear and buoyancy on the flow structure and heat and momentum transfer, that is $(Nu_h, Nu_\omega)=f(Ra,\Omega)$. Therefore, referring to the results of our previous experiments and simulations of ACRBC \citep{jiang_supergravitational_2020,jiang_experimental_2022,wang_spectra_2022}, $Pr=4.3$ is taken for water at $40^{\circ}C$, and $\eta$ is set to be $0.5$. Considering the Boussinesq approximation, $\alpha\Delta\ll 1$ is required. According to equations (\ref{ParEqu}), we choose $\alpha\Delta=6.67\times 10^{-3}$ and then $Ro^{-1}=20$, which is also in the parameter range of our previous experiments, corresponding to a temperature difference $\Delta\approx17.3K$ for water \citep{jiang_supergravitational_2020,jiang_experimental_2022}. The Non-Oberbeck-Boussinesq effect is not obvious at this temperature difference \citep{ahlers_2006} and the Oberbeck-Boussinesq conditions are well satisfied. Meanwhile, considering the definition of $Ro^{-1}$, in the non-dimensional system, the rotational speed of the rotating reference frame is $(\Omega_cL/U)=Ro^{-1}/2=10$. 

As for instability, we focus on $Ra\in[10^3,10^9]$ and $\Omega\in[10^{-2},10]$, where positive $\Omega$ means the inner cylinder rotates at a faster angular velocity than the outer cylinder in the stationary reference frame. For example, $\Omega=10$ corresponds to the state that the inner cylinder is rotating with the non-dimensional angular velocity $\omega =20$ and the outer cylinder is static in the stationary reference frame. The DNS for heat tranfer analysis covers a $Ra$ range $[10^6,10^8]$ and a rotational speed difference $\Omega$ range $[10^{-2},10]$. 

\section{Results and discussion}\label{sec3}

Previous studies have discussed or implied how the flow and heat transfer are at both ends of our parameter range. When $\Omega=0$, Jiang {\it et al.} \citep{jiang_supergravitational_2020} shows that the flow will be quasi-two-dimensional at high $Ro^{-1}$ number ($Ro^{-1}\ge10$), due to the constraint of Taylor-Proudman theorem. The scaling law of heat transfer in this region follows the Grossman-Lhose (GL) theory, with a power-law relationship of $Nu\sim Ra^{0.27}$ in the classical regime. At the other end, $\Omega=10$, only the inner cylinder rotating, previous work \citep{leng_flow_2021,leng_mutual_2022} implies that the shear is so strong that the flow is dominated by the TC vortex and the temperature behaves like a passive scalar. In this section, we will reveal how the flow undergoes such a large transformation, from the RB flow in the $r\varphi$ plane to the TC flow developed in the $rz$ plane.

\subsection{Flow regimes}

\begin{figure}
	\centering
	\includegraphics[width=0.8\linewidth]{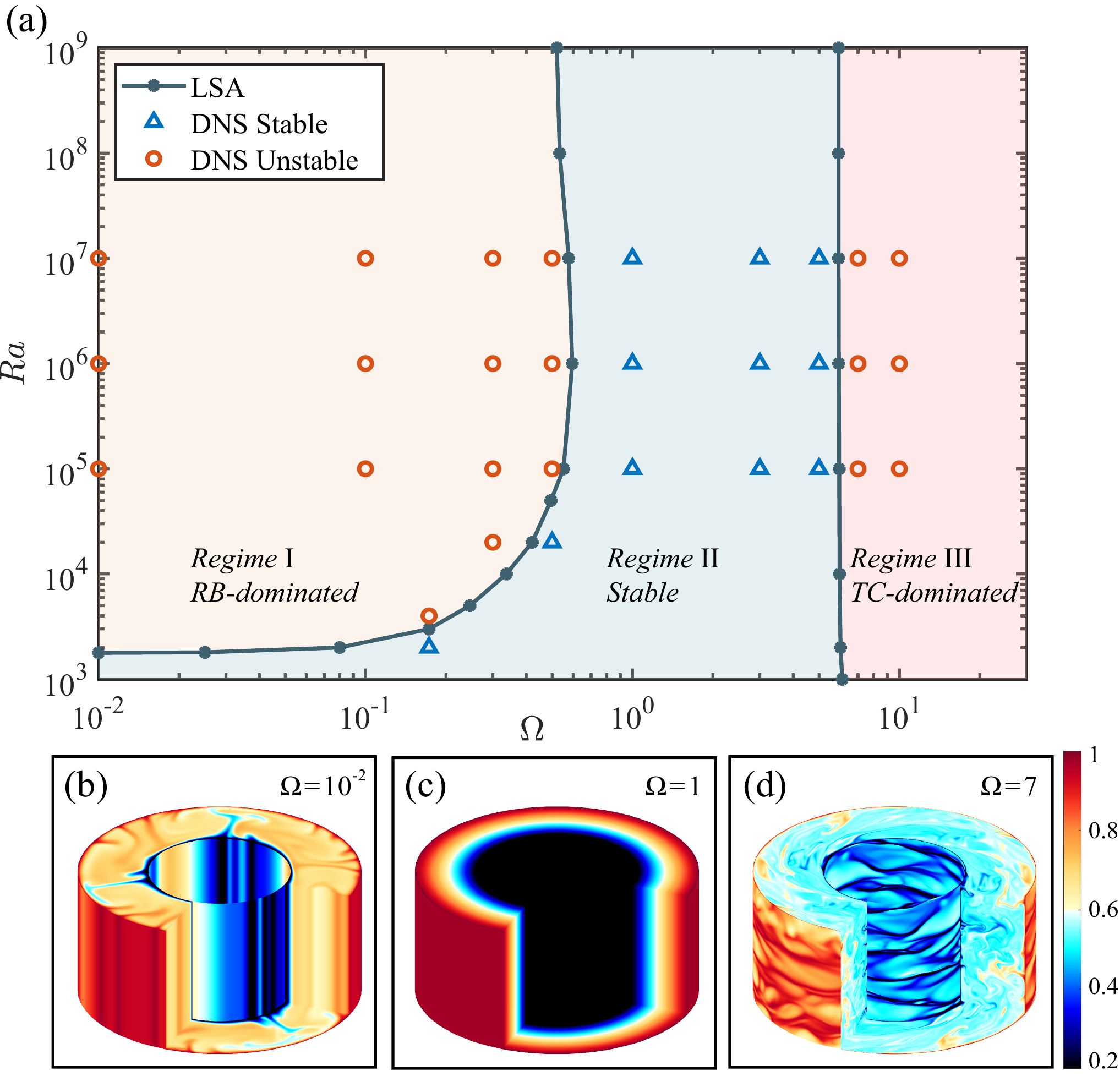}
        \captionsetup{justification=raggedright}
	\caption{(\textit{a}) Instability regimes divided by LSA (the bottle green lines) and checked by DNS (the red circles and the blue triangles) in the $(Ra,\Omega)$ parameter domain. (\textit{b}) Instantaneous temperature fields from DNS at $Ra=10^6, \Omega=10^{-2}$ (Regime I). (\textit{c}) Instantaneous temperature fields from DNS at $Ra=10^6, \Omega=1$ (Regime II). (\textit{d}) Instantaneous temperature fields from DNS at $Ra=10^6, \Omega=7$ (Regime III).}
	\label{fig:Regimes}
\end{figure}

When increasing shear converts the flow from the RB flow to the TC flow, a stable regime is found on the way. To determine the boundaries of the stable regime, the approach of linear stability analysis is applied and the results are checked by the direct numerical simulation. As shown in figure \ref{fig:Regimes}(a), two unstable regimes are distributed on the two sides of the parameter domain and a stable regime locates in the middle. The results of DNS agree well with the boundaries given by LSA, guaranteeing the validity of the stable regime. We denote the three regimes as RB-dominated regime (Regime I), stable regime (Regime II), and TC-dominated regime (Regime III), with the shear strength increasing. 

In the stable regime (Regime II), the typical instantaneous temperature field is illustrated in figure \ref{fig:Regimes}(c). The system is governed by steady laminar and nonvortical TC flow, as described by equations (\ref{LamEqu}). In the other two regimes, the typical instantaneous temperature fields exhibit significant differences, depicted in figures \ref{fig:Regimes}(b,d). In Regime I, the flow is quasi-two-dimensional on the $r\varphi$ plane. Plumes are detached from the boundary layers, move across the bulk region, and transfer heat to the other side. Several convection roll pairs are generated, the number depending on the radius ratio $\eta$ when there is no shear \citep{pitz_onset_2017,wang_effects_2022}. In Regime III, the flow is three-dimensional, and the roll pairs, or called Taylor vortexes, are mainly on the $rz$ plane, which can be seen clearly on the vertical $\varphi$ slice in figure \ref{fig:Regimes}(c). There are large differences among the flow structures in these three regimes. For a fixed $Ra$, as the shear increases from zero, the flow first stabilizes from a quasi-two-dimensional RB flow into a laminar flow and then destabilizes into a three-dimensional flow similar to the TC flow. It's worth noting that as $\eta=0.5$ in our system, the Richardson number $Ri=\Omega^{-2}$, based on the equation (\ref{RiEqu}). Therefore, at high $Ra$, the stable regime locates at $Ri\sim O(1)$, which means the buoyancy is in the same order with shear; this is consistent with the transition region from buoyancy-dominated to shear-dominated obtained from previous work \citep{blass_flow_2020,leng_flow_2021}.

The curves meaning the marginal state between regimes contain important information. The boundary between Regimes I and II has a horizontal asymptote on the left side for $\Omega\rightarrow 0$, that is the onset $Ra$ of the ACRBC at $\eta=0.5$. As $\Omega$ increases, the instability of the flow is suppressed and a stronger buoyancy force is required to drive the convection. The critical Rayleigh number $Ra_c$ of the marginal state grows faster and faster with increasing $\Omega$, and finally, when $\Omega$ exceeds a certain value (about $0.6$ with our parameter settings), the flow is always stable no matter how much the Rayleigh number increases in our parameter range. According to the marginal-state curve in figure \ref{fig:Regimes}(a), there may also exist a vertical asymptote for $Ra\rightarrow\infty$. With the participation of the shear, the flow instability problem is different from the onset of the RB convection. For the onset $Ra$ of the ACRBC without shear, there exists a perturbation energy balance between the work performed by the buoyancy and the energy dissipation by the fluid viscosity; when the shear is imposed, the movement of the boundaries changes the balance.
In addition, we notice that for the sheared classical rectangular RB system, the flow does not become stable with increasing shear \citep{blass_flow_2020}. This indicates that the effect of curvature and the rotation in sheared ACRBC may also contribute to the stabilization of the flow. We will discuss the physics of the instability between Regime I and II in depth in section \ref{sec:instability}.

Moreover, the straight boundary between Regimes II and III seems to behave differently. On this boundary, $Ta$ is much larger than $Ra$: according to the equation (\ref{TaEqu}), when $Ra=10^3$, $Ta\approx 1\times10^5$. Therefore, the marginal-state curve between Regime II and III is similar to the line given by Rayleigh's inviscid criterion \citep{ali_stability_1990,drazin_reid_2004,yoshikawa_instability_2013}, with the thermal effect enhancing the instability slightly. As the instability of TC flow with radial temperature difference has been widely discussed \citep{ali_stability_1990,yoshikawa_instability_2013,kang_thermal_2015,meyer_effect_2015,yoshikawa_linear_2015}, we mainly focus on the instability in Regime I and II in the following discussion.

\subsection{Instability}\label{sec:instability}

\subsubsection*{Mode analysis}

To investigate the effect of shear on the flow and how the stable regime arises, we begin with linear instability. In Regime I, the flow is unstable, and each pair of azimuthal and axial wavenumber $(n,k)$ is associated with a perturbation mode of a growth rate $\sigma$. We notice that for every $n$, $\sigma$ reaches a maximum all at $k=0$, corresponding to the quasi-2D flow characteristic in Regime I, that the flow is limited on the $r\varphi$ plane. Therefore, only $k=0$ is considered in subsequent analysis on Regime I.

\begin{figure}\label{Mode_n}
	\centering
	\includegraphics[width=0.9\linewidth]{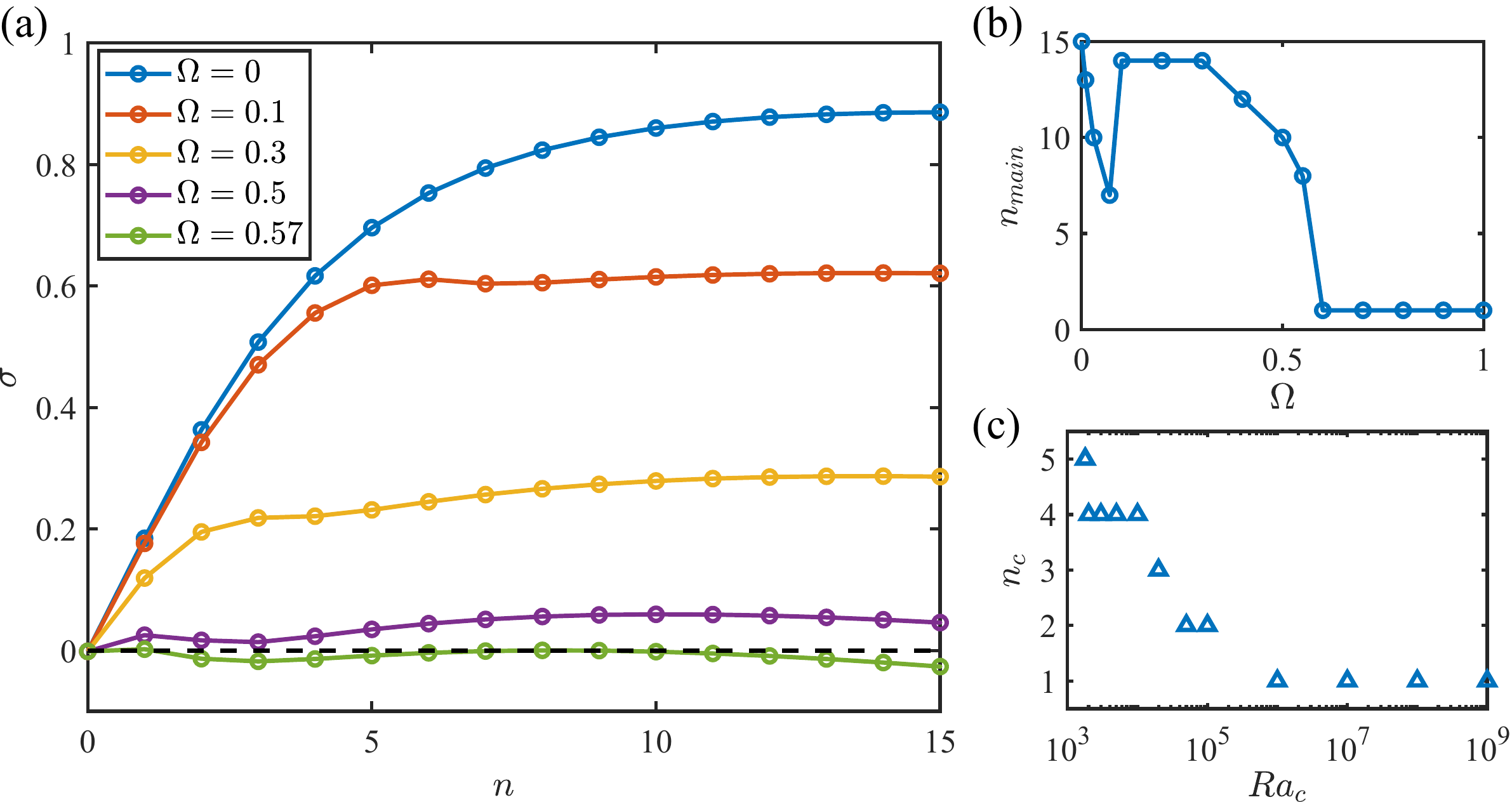}
        \captionsetup{justification=raggedright}
	\caption{(\textit{a}) Growth rate $\sigma$ as a function of the azimuthal wavenumber $n$ for $\Omega=0, 0.1, 0.3, 0.5$ and $0.57$., at $Ra=10^7$. The black dashed line denotes $\sigma=0$. (\textit{b}) The azimuthal wavenumber of the main mode $n_{main}$ as a function of $\Omega$, at $Ra=10^7$. (\textit{c}) The critical azimuthal wavenumber $n_c$ as a function of the critical Rayleigh number $Ra_c$ at the marginal state.}
	\label{fig:Modes_n}
\end{figure}

Wondering how shear influences the flow instability, we analyze the growth rate $\sigma$ as a function of the azimuthal wavenumber $n$ for increasing $\Omega$ at $Ra=10^7$, and the results are plotted in figure \ref{fig:Modes_n}(a). In the no shear case $\Omega=0$, $\sigma$ grows with $n$ quickly at first and then reaches a plateau. Modes of high azimuthal frequency $n\gtrsim 5$ are preferred, having a higher growth rate than the low azimuthal wavenumber modes. With the shear imposed, the growth rates of all modes decrease synchronously, while the high-frequency modes are suppressed more. This means that shear suppresses the growth of linear instability and further influences the generation of the convection flow. Meanwhile, perturbations with higher azimuthal frequency modes are suppressed more, and the dominated mode becomes the low-frequency mode gradually, as an increasing shear is imposed. Figure \ref{fig:Modes_n}(b) demonstrates this trend. The azimuthal wavenumber of the main mode (the mode owning maximum growth rate $\sigma$) $n_{main}$ decreases from $15$ to $1$ with the increasing shear. When the flow approaches the marginal state, $n_{main}$ drops to $1$ and holds on. Figure \ref{fig:Modes_n}(c) shows at the marginal state, how the critical azimuthal number $n_c$ changes with the critical Rayleigh number $Ra_c$. At no shear state, which is the left asymptote in figure \ref{fig:Regimes}(a), $n_c=5$, giving the mode of five roll pairs. As the Rayleigh number and the shear increase, $n_c$ drops from $5$ to $1$, and holds on at high Rayleigh number $(Ra_c\ge 10^6)$. It's worth noting that in figure \ref{fig:Regimes}(a), the critical rotating difference $\Omega_{cr}$ reaches a maximum near $Ra=10^6$ as well, and then the $\Omega_{cr}$ required to stabilize the flow decreases slightly with increasing $Ra$. The slight decrease of $\Omega_{cr}$ may be related to the fact that the critical azimuthal wavenumber reaches the minimum.

\begin{figure}
	\centering
	\includegraphics[width=1.0\linewidth]{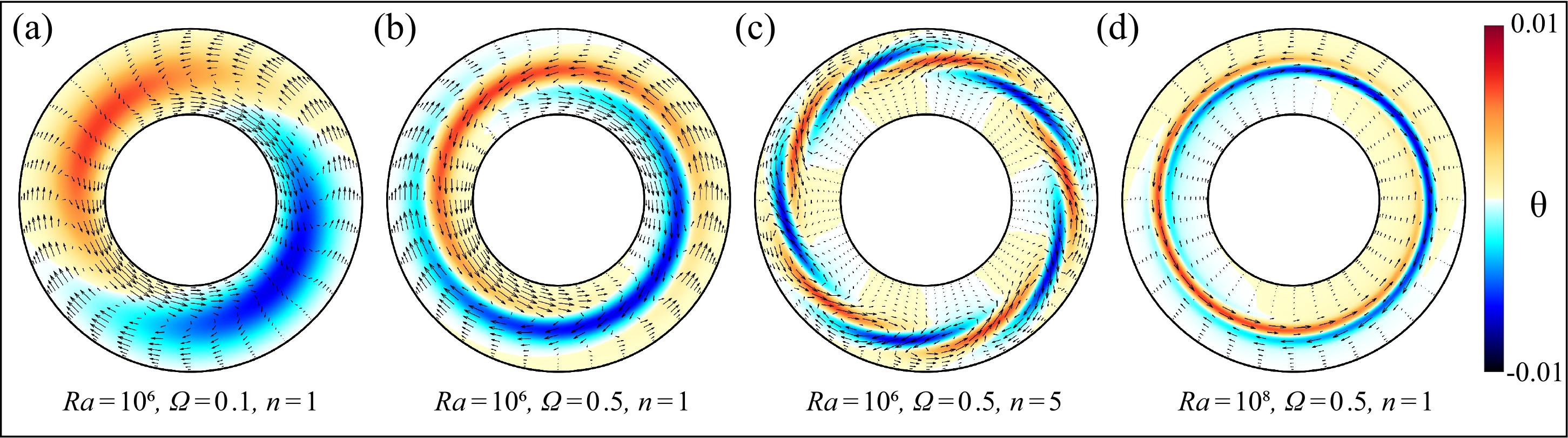}
        \captionsetup{justification=raggedright}
	\caption {Eigenfunctions $(\boldsymbol{u'}, \theta')$ of the unstable modes at corresponding conditions: (\textit{a}) $Ra=10^6,\Omega=0.1,n=1$, (\textit{b}) $Ra=10^6,\Omega=0.5,n=1$, (\textit{c}) $Ra=10^6,\Omega=0.5,n=5$, (\textit{d}) $Ra=10^8,\Omega=0.5,n=1$. The maximum temperature perturbation is set to be the same in all conditions, $|\theta'|_{max}=10^{-2}$. The velocity perturbation vectors are scaled by the maximum velocity magnitude of each, as $|\boldsymbol{u'}|_{max}$ equals to (\textit{a}) $7.3\times 10^{-3}$, (\textit{b}) $4.0\times 10^{-3}$, (\textit{c}) $4.2\times 10^{-3}$, (\textit{d}) $4.0\times 10^{-3}$.}
	\label{fig:Modes_TU}
\end{figure}

The temperature and the velocity eigenfunctions of certain modes normalized by the maximum temperature perturbations are illustrated in figure \ref{fig:Modes_TU}. For $n=1$, there is only one hot-cold perturbation pair, while for $n=5$ there are five pairs. In all perturbation modes, the colder flow goes outward from the inner side with a clockwise azimuthal velocity, in the opposite direction of the basic flow. When the system is under low shear, as shown in figure \ref{fig:Modes_TU}(a), the azimuthal motion of the flow is strong at the junction of the cold and hot flow, while relatively weak at the center of cold and hot perturbations. Near the inner wall, two flows with opposing azimuthal velocity converge at the center region of cold temperature perturbations. As a result of the mass conservation, the flow undergoes an outward deflection, thereby transporting the strong cold temperature perturbations toward the outer wall. Similar structures exist extensively in the modes under various conditions, as illustrated in figures \ref{fig:Modes_TU}(b-d). Comparing figure \ref{fig:Modes_TU}(b) to (a), as the shear becomes stronger, we notice that the cold and hot perturbations are more stretched azimuthally and are narrower radially, which is similar to the plumes in sheared RB convection \citep{goluskin_convectively_2014,blass_flow_2020}. In figure \ref{fig:Modes_TU}(b), a strong azimuthal velocity exists everywhere in the cold and hot flow, even in the center region of the cold flow, different from the low-shear case. The main motion of the flow is in the azimuthal direction, which has a negative effect on heat transfer and the work done by buoyancy. Under stronger shear, it seems more difficult to transport heat from one side to the other side. Both larger dissipation and less work done by buoyancy depress the growth of the instability.

Moreover, for the modes of high azimuthal frequency under strong shear, more pairs occupy the azimuthal length of $2\pi$, an example of which is shown in figure \ref{fig:Modes_TU}(c). The temperature perturbation pairs and flow roll pairs are denser, presenting a pattern that extends counterclockwise from the inside out. As can be expected, this mode has higher dissipation and more significant buoyancy work than the low azimuthal frequency mode $n=1$ in figure \ref{fig:Modes_TU}(b). In addition, we are interested in the instability case when moving to high Rayleigh numbers. A mode near the marginal state at $Ra=10^8$ is presented in figure \ref{fig:Modes_TU}(d). Interestingly, the perturbations exhibit a pronounced concentration within an exceedingly narrow annulus, approximately located at the intermediate radius of the system. Inside and outside of this annulus, both temperature and velocity perturbations are weak. As $Ra$ escalates, the annulus progressively diminishes in width, tending to converge to a certain radius, which may be associated with the limit of $\Omega_{cr}$ for an infinitely-great Rayleigh number.

\subsubsection*{Energy analysis}

\begin{figure}
	\centering
	\includegraphics[width=1.0\linewidth]{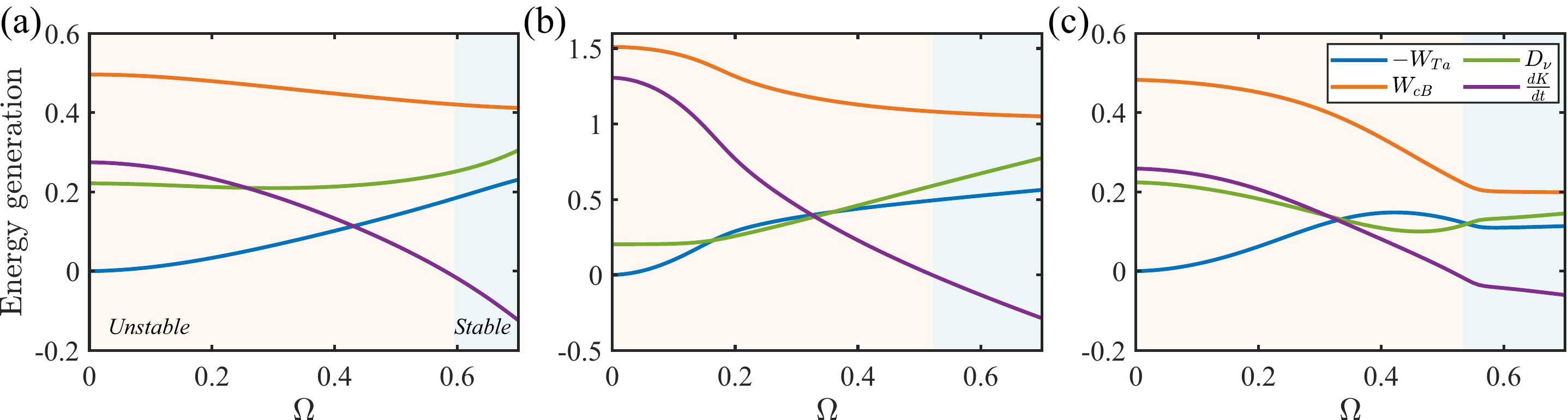}
        \captionsetup{justification=raggedright}
	\caption {Variation of energy generation terms $-W_{Ta}$,$D_\nu$,$W_{cB}$ and the growth rate of kinetic energy $\frac{dK}{dt}$ with $\Omega$, at conditions (\textit{a}) $Ra=10^6,n=1$, (\textit{b}) $Ra=10^6,n=5$, (\textit{c}) $Ra=10^8,n=1$. All terms are normalized by $K$. The modes in the light orange region (left side) are unstable, while in the nattier blue region (right side) are stable.}
	\label{fig:Energy}
\end{figure}

In order to get a comprehensive explanation of the instability in Regime I, we perform the energy analysis on perturbations. Multiply the velocity perturbations $u'_r,u'_\varphi,u'_z$ by the linearized momentum equations in their respective directions and take their summation, then the kinetic energy equation of perturbations is obtained \citep{yoshikawa_instability_2013,yoshikawa_linear_2015,meyer_effect_2015}:
\begin{equation}\label{EnergyEqu}
    \frac{dK}{dt}=W_{Ta}+W_{cB}-D_\nu,
\end{equation}
	where the kinetic energy $K$, the power of  inertial forces $W_{Ta}$, the power of centrifugal buoyancy $W_{cB}$ and the dissipation $D_\nu=\langle\Phi\rangle$ are expressed as: 
\begin{equation}\label{TermsEqu}
    \begin{aligned}
        K=\frac{1}{2}\langle|\boldsymbol{u'}|^2\rangle,\ 
        W_{Ta}&=\langle-u'_ru'_\varphi(\frac{dV}{dr}-\frac{V}{r})\rangle,\\
        W_{cB}=\langle-\frac{2(1-\eta)ru'_r}{1+\eta}[\theta'(1&+\frac{2V}{Ro^{-1}r})^2+\frac{4\Theta u'_\varphi}{Ro^{-1}r}(1+\frac{2V}{Ro^{-1}r})] \rangle,\\
        D_\nu=\sqrt{\frac{Pr}{Ra}}\langle 2[(\frac{\partial u'_r}{\partial r})^2&+(\frac{1}{r}\frac{\partial u'_\varphi}{\partial{r}}+\frac{u'_r}{r})^2+(\frac{\partial u'_z}{\partial z})^2]+[r\frac{\partial}{\partial r}(\frac{u'_\varphi}{r})+\frac{1}{r}\frac{\partial u'_r}{\partial\varphi}]^2\\
        &+[\frac{1}{r}\frac{\partial u'_z}{\partial\varphi}+\frac{\partial u'_\varphi}{\partial z}]^2+[\frac{\partial u'_r}{\partial z}+\frac{\partial u'_z}{\partial r}]^2\rangle.\\
    \end{aligned}
\end{equation}
The angle brackets $\langle\cdot\rangle$ denote the average over the whole space $r,\varphi$, and $z$. A mode is stable if the right-hand side of the equation (\ref{EnergyEqu}) is negative. The kinetic energy generation terms $W_{Ta},W_{cB},D_\nu$ normalized by the kinetic energy $K$ are illustrated in figure \ref{fig:Energy} as functions of $\Omega$ at $Ra=10^6$ and $10^8$. As all terms are normalized by the averaged kinetic energy $K$, the left side of the equation (\ref{EnergyEqu}) $dK/dt$ is twice the growth rate, derived from equation (\ref{equ：decomp}). As shown in figure \ref{fig:Energy}(a), $Ra=10^6$ and $n=1$, with $\Omega$ increasing, the power of centrifugal buoyancy $W_{cB}$ decreases and the dissipation term $D_\nu$ increases slightly, in line with our analysis of the modes in figure \ref{fig:Modes_TU}. The largest change is in the inertial force action term, $W_{Ta}$, which is zero under no shear but its magnitude increases rapidly as $\Omega$ increases. According to the formula of $W_{Ta}$ in equations (\ref{TermsEqu}), this term is closely related to the shear action. When the system reaches the marginal state, three energy generation terms exhibit equilibrium. The presence of $W_{Ta}$ constitutes the primary differentiating factor between the flow instability with shear and the flow instability without shear.

The energy analysis method can also explain the reasons behind the greater suppression of high azimuthal frequency modes in comparison to low-frequency modes to a certain degree. Figure \ref{fig:Energy}(b) shows the variation of energy generation terms under $Ra=10^6, n=5$. At $\Omega=0$, the centrifugal buoyancy term $W_{cB}$ is much larger than dissipation, injecting a lot of energy into the development of the instability. However, with shear imposed, both $-W_{Ta}$ and $D_\nu$ rise quickly, while the buoyancy term $W_{cB}$ drops faster than in the low-frequency mode $n=1$. This reflects the strong suppression of the high-frequency modes by shear. Unlike the low-frequency mode in figure \ref{fig:Energy}(a), the dissipation term $D_\nu$ plays a more important role, increasing significantly with shear, which is closely related to the tight distribution of velocity perturbations rolls. Therefore, although the high-frequency mode offers a high growth rate at no shear, it reaches the marginal state earlier as the shear increases. 

At high $Ra$, the dissipation term decreases at first with increasing shear, as illustrated in figure \ref{fig:Energy}(c). The buoyancy terms drop more than the mode at $Ra=10^6$, as the perturbations are concentrated in a narrow annulus and the azimuthal motion occupies more kinetic energy. Meanwhile, we notice that at the marginal state, it is still a balance of three items $W_{Ta}, W_{cB}, {D_\nu}$, which can be further observed at higher $Ra=10^{13}$. Therefore, the limit of instability as the Rayleigh number tends to infinity can not be explained by an inviscid solution, since the dissipation term $D_\nu$ is not negligible.

\begin{figure}
	\centering
	\includegraphics[width=1.0\linewidth]{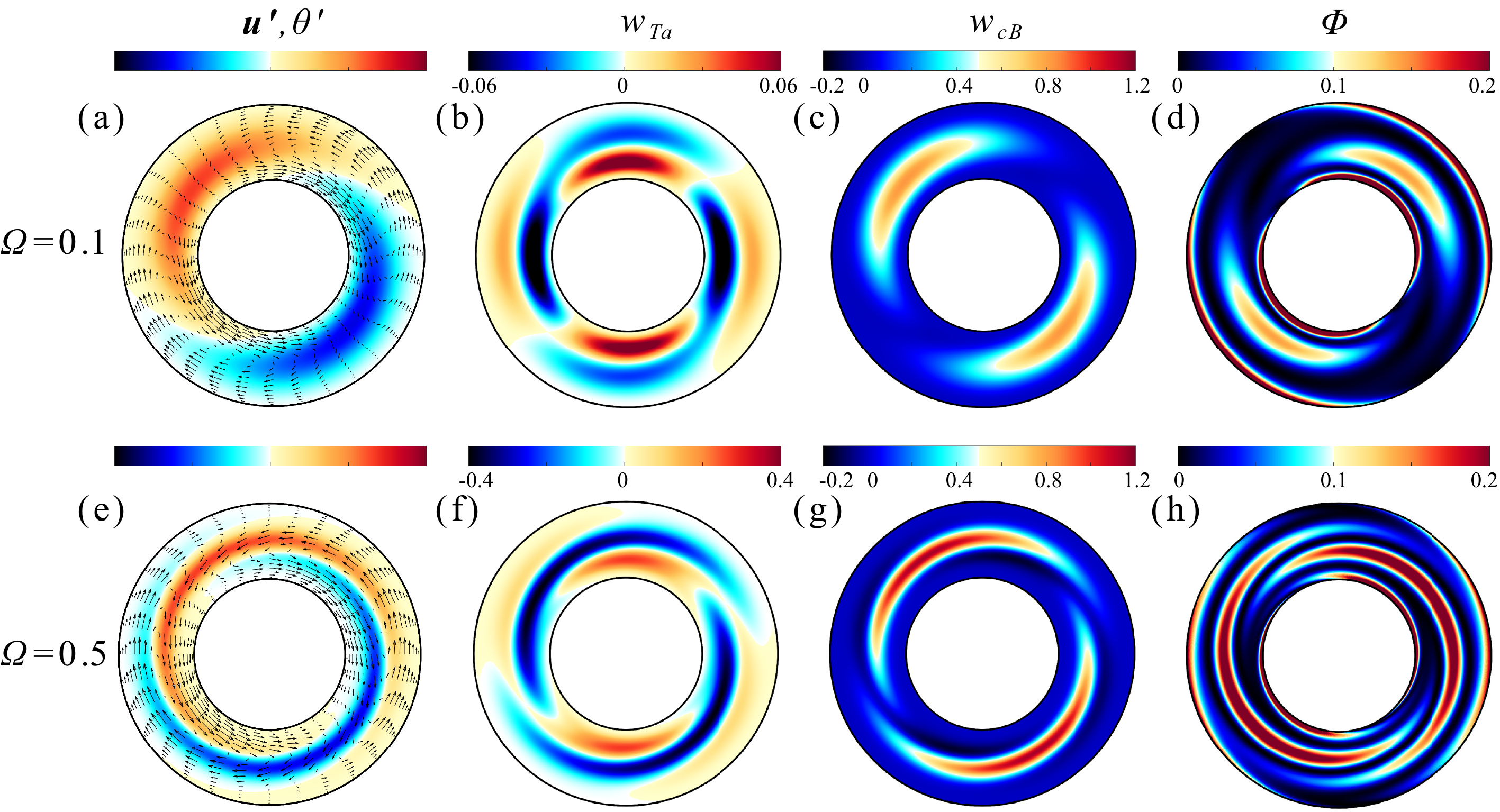}
        \captionsetup{justification=raggedright}
	\caption {Distribution of perturbations (\textit{a,e}) $(\boldsymbol{u'},\theta')$ and the densities of energy generation terms (\textit{b,f}) $w_{Ta}$, (\textit{c,g}) $w_{cB}$ and (\textit{d,h}) $\Phi$ under different shear (\textit{a-d}) $\Omega=0.1$, (\textit{e-f}) $\Omega$=0.5 when $Ra=10^6, n=1$. All the densities of energy generation terms are normalized by the averaged kinetic energy $K$. Note that the scales of $w_{Ta}$ in (b) and (f) are different.}
	\label{fig:Energymodes}
\end{figure}

Moreover, at $Ra=10^6$, the distribution of energy generation terms, $w_{Ta}, w_{cB}, \Phi$ are compared under different shear strengths with the azimuthal wavenumber $n=1$, as shown in figure \ref{fig:Energymodes}. The densities of energy generation terms are also normalized by $K$. By comparing figure \ref{fig:Energymodes}(b) and figure \ref{fig:Energymodes}(f), one can find that the power of inertial term $w_{Ta}$ increases substantially with the enhancement of shear. (Note: the scales of the color bars in these two figures are different.) Negative $w_{Ta}$ is derived from the rightward deflection of outward or inward flow ($u'_ru'_\varphi<0$); while the positive part mainly comes from the limit of mass conservation near the boundary and occurs at the junction of cold and hot currents. As for the buoyancy power $w_{cB}$, the increase of shear amplifies its maximum intensity while making it more concentrated in distribution, as the cold and hot perturbations are more stretched azimuthally. Dissipation term $\Phi$ mainly occurs near the boundary and the regions with opposing flows of cold and hot currents, where a strong velocity gradient exists. As shear increases, $\Phi$ increases significantly at the region where two flows meet, contributing to a final increase of $D_\nu$. These qualitative interpretations are in relatively good agreement with the curves in figure \ref{fig:Energy}(a).

\subsection{Heat and momentum transfer}\label{heatsec}

\begin{figure}
	\centering
	\includegraphics[width=1.0\linewidth]{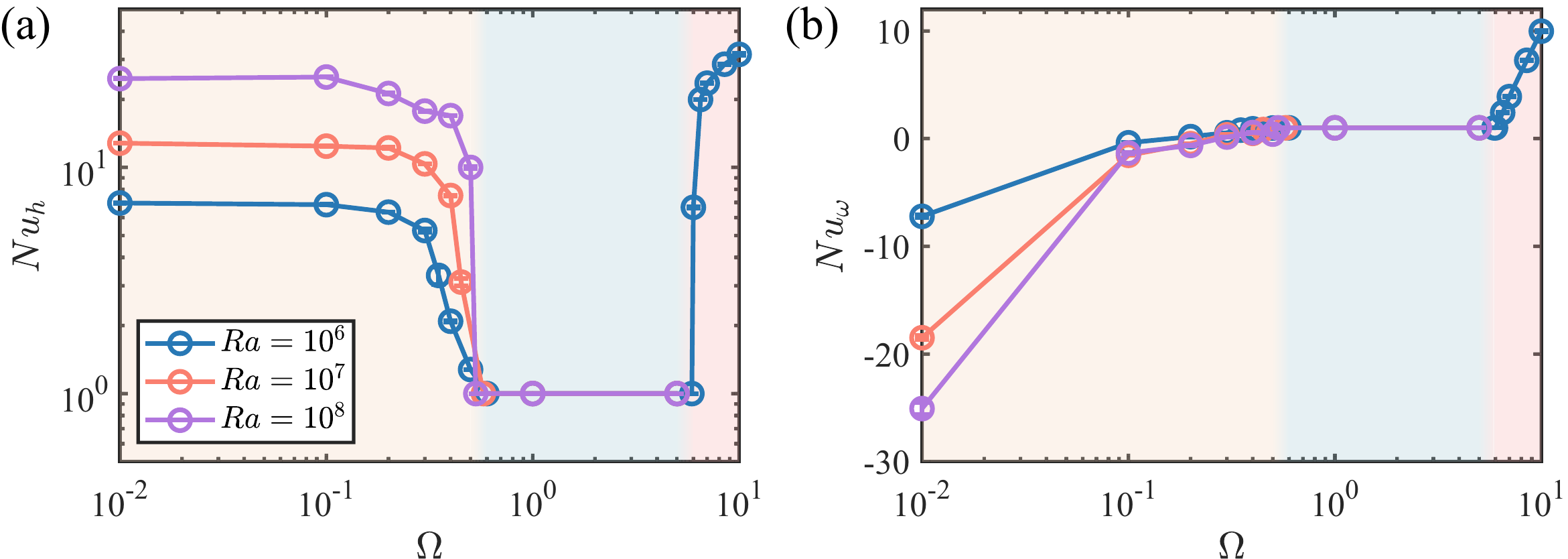}
	\caption {Variation of (\textit{a}) $Nu_h$ and (\textit{b}) $Nu_\omega$ with $\Omega$, at $Ra=10^6$, $10^7$ and $10^8$.}
	\label{fig:Nu}
\end{figure}

In this subsection, we focus on both heat and momentum transport efficiency and discuss their behaviors in different regimes. Under the dimensionless formulation, the heat transfer efficiency and the momentum transfer efficiency are measured by two Nusselt numbers: $Nu_h$ and $Nu_\omega$, defined as the ratios of the corresponding currents of the system to the currents in the laminar and nonvortical flow case \citep{eckhardt_torque_2007,wang_effects_2022}:
\begin{equation}
	\begin{aligned}
		Nu_h&=\frac{\sqrt{RaPr}\langle u_r\theta\rangle_{t,\varphi,z}-\partial_r\langle\theta\rangle_{t,\varphi,z}}{(rln(\eta))^{-1}},\\
		Nu_\omega&=\frac{r^3[Ra/Pr\langle u_r\omega\rangle_{t,\varphi,z}-\sqrt{Ra/Pr}\partial_r\langle\omega\rangle_{t,\varphi,z}]}{2B},\\
	\end{aligned}
\end{equation} 
where $\omega=u_\varphi/r$ is the angular velocity of the fluid, and $B$ is the parameter of the base flow defined in equations (\ref{LamEqu}). The variation of the two Nusselt numbers with the shear strength $\Omega$ under different Rayleigh numbers are illustrated in figure \ref{fig:Nu}. The different regimes in the figure are distinguished using different background colors. As the Taylor number is much larger than $Ra$ in Regime III ($Ta=2.65\times 10^9$ at $Ra=10^7, \Omega=10$, calculated from the equation (\ref{TaEqu})), the flow is too drastic and computationally expensive, so we only calculate the cases of $Ra = 10^6$ in the TC-dominated regime. 

In Regime I, as the imposed shear increases, $Nu_h$ decreases slowly at first and then rapidly when $\Omega>0.3$. The curves of different $Ra$ behave in a similar trend. The heat transfer efficiency $Nu_h$ increases with $Ra$ in Regime I, but as the shear increases and flow turn stable, $Nu_h$ under different $Ra$ all drop to $1$. These results are similar to the results in the sheared rectangular RB cell \citep{blass_flow_2020}, where the heat transfer is also first depressed by shear. However, in sheared rectangular RB cell, shear can not turn the flow laminar. While for momentum transfer, we notice that the system gives a negative $Nu_\omega$ under a weak shear, which means the large-scale circulation (LSC) in the RB convection pushes the two cylinders to rotate relative to each other. According to the definition of $Nu_\omega$, when there is no relative rotation between two cylinders, LSC also imposes shear stress to the wall, and $Nu_\omega$ is negative infinity. $Nu_\omega$ increases quickly as $\Omega$ increases, and the growth trend slows down when its value is close to $1$. Similar to the heat transfer, a larger $Ra$ gives a stronger torque. 

\begin{figure}
	\centering
	\includegraphics[width=1.0\linewidth]{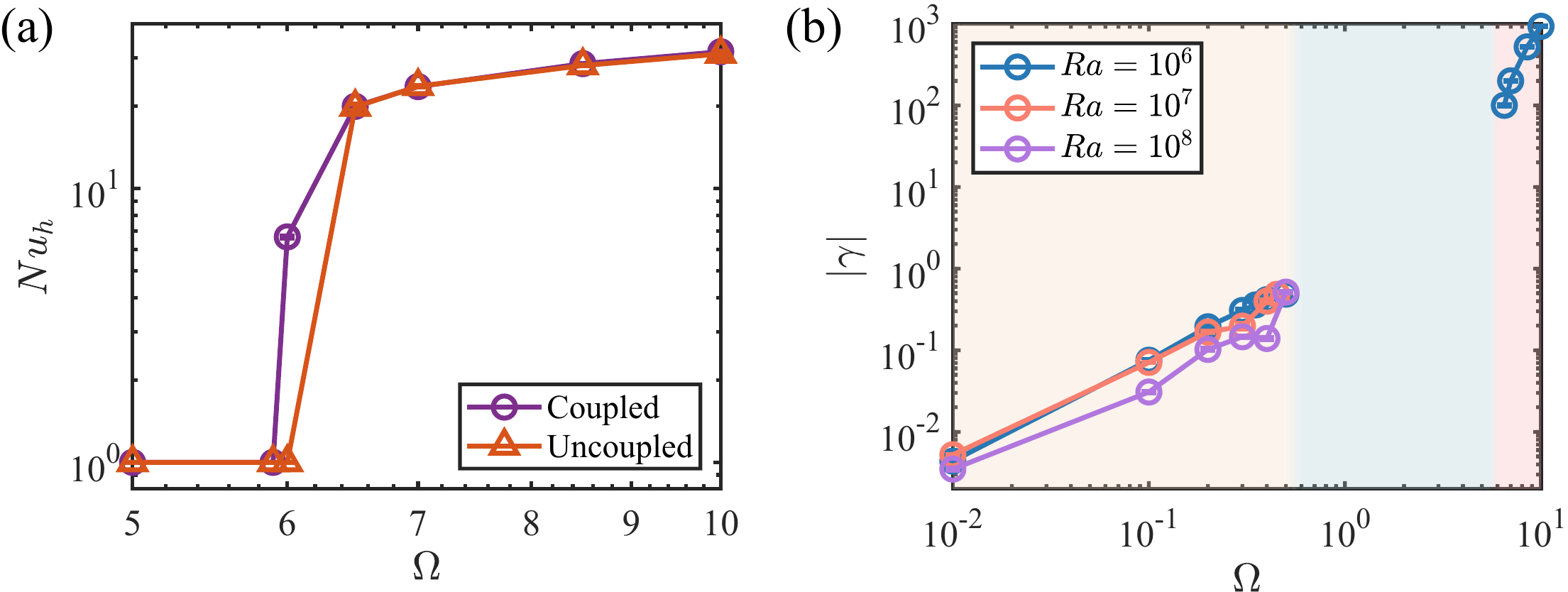}
        \captionsetup{justification=raggedright}
	\caption {(\textit{a}) Variation of $Nu_h$ with $\Omega$ in Regime III, for the cases that the temperature and velocity are coupled (the purple line) and uncoupled (the red line). $Ra=10^6$. (\textit{b}) Variation of $|\gamma|$ with $\Omega$ at $Ra=10^6, 10^7, 10^8$. $\gamma<0$ in Regime I and $\gamma>0$ in Regime III.}
	\label{fig:Nu2}
\end{figure}

In Regime II, the flow field is laminar, described by equations (\ref{LamEqu}). Both $Nu_h$ and $Nu_\omega$ equal to $1$. As $\Omega$ increases, the system reaches the marginal state between Regime II and III. If the buoyancy is not considered, at high $Ta$, the marginal-state curve of Taylor instability is close to the solution given by Rayleigh inviscid theory \citep{drazin_reid_2004}, which gives $\Omega=6$. Considering the buoyancy, the instability is strengthened and the flow at $\Omega=6$ gives a heat transfer efficiency close to $Nu_h$ under weak shear. Meanwhile, $Nu_\omega$ remains almost unchanged, which means the flow driven by buoyancy is hard to mix the angular velocity profile. As the shear strength continues to increase, the heat transfer is enhanced more, and the momentum transfer increases gradually as well. It is suggested that this regime appears in a similar system with gravity-driven buoyancy, where the system is mainly controlled by the Taylor vortex and temperature acts as a passive scalar \citep{leng_flow_2021}. To check if this idea is also valid in our system, we perform another set of simulations under the same conditions, where the velocity is decoupled from temperature (buoyancy not considered) and the temperature really acts as a passive scalar. The comparison of the two is displayed in figure \ref{fig:Nu2}(a). As there is no buoyancy, the system is still stable at $\Omega=6$, but then it behaves almost the same as the velocity-temperature-coupled system, at $\Omega=6.5$ in the figure. Therefore, in most of Regime III, the temperature can be regarded as a passive scalar in our system; it is the shear that dominates the flow.

The exact relation gives us an interesting perspective to understand the relation between $Nu_h$ and $Nu_\omega$. Derivated by the kinetic energy balance equation, the exact relation represents the energy balance between buoyant energy injection, shear energy injection, and mechanical dissipation. In the dimensional form, referring to derivation from Eckhardt, Grossmann, and Lohse \citep{eckhardt_torque_2007} and Wang {\it et al.} \citep{wang_spectra_2022}, if we assume $\langle\alpha\theta(\Omega_c+u_\varphi/r)^2ru_r\rangle_{V,t}=C\alpha\Omega_c^2\langle ru_r\theta\rangle_{V,t}$, the exact relation of our system is:
\begin{equation}\label{NuEqu}
    \varepsilon-\varepsilon_{lam}=\nu^{3}L^{-4}[\sigma_r^{-2}Ta(Nu_\omega-1)+Cf(\eta)Pr^{-2}Ra(Nu_h-1)],
\end{equation}
where $\varepsilon=\nu\langle(\partial_iu_j+\partial_ju_i)^2\rangle_{V,t}$ is the mean energy dissipation rate, $\varepsilon_{lam}$ is the mean energy dissipation rate of the laminar and nonvortical flow, $\sigma_r=(1+\eta)^4/16\eta^2$ is the quasi-Prandtl number, and $f(\eta)=\frac{2(\eta-1)}{(1+\eta)ln(\eta)}$ is a correction factor. In Regime I, $u_\varphi/r\ll\Omega_c$, then $C\approx 1$. In Regime III, the flow is dominated by shear; therefore, referring to the angular velocity profile of the TC flow \citep{grossmann_highreynolds_2016}, $C$ is still at the order $O(1)$. 

The terms on the right of the equation (\ref{NuEqu}) represent the energy injection by the wall shear and the buoyancy, respectively. The TC flow considers the first term while the RB convection concentrates on the second. In Regime I, as the $Nu_h>1$ and $Nu_\omega<1$, only buoyancy provides energy for turbulent dissipation, while the wall motion consumes energy instead. In Regime III, buoyancy and shear provide energy to the fluid together. Moreover, considering the ratio of the two terms, we can verify the dominant regimes of the flow. The ratio $\gamma$ can be simplified by take $\eta=0.5$, $Pr=4.3$ and $C\approx 1$:
\begin{equation}
	\gamma=\frac{\sigma_r^{-2}Ta(Nu_\omega-1)}{Cf(\eta)Pr^{-2}Ra(Nu_h-1)}\approx31.79\Omega^2\frac{Nu_\omega-1}{Nu_h-1},
\end{equation}
where $Ta/Ra$ is simplified using the equation (\ref{TaEqu}). How $\gamma$ varies with $\Omega$ is shown in figure \ref{fig:Nu2}(b). In Regime I, $\gamma<0$; in Regime II, $\gamma$ is not calculated as both terms are zero; in Regime III, $\gamma>0$. When the system is under weak shear, such as $\Omega=10^{-2}$, a very small fraction of the energy is consumed by shearing. As $\Omega$ increases, more energy is transferred out of the system by the wall motion, as $|\gamma|$ grows approximately linearly with $\Omega$ in logarithmic coordinates. At the point near Regime II, $\Omega=0.5$, about half of the energy injected by buoyancy is consumed by the wall shear. Such a large percentage of energy consumption may be responsible for bringing the system to a stable state. Crossing Regime II into Regime III, $\gamma\gg 1$, the energy input by the wall shear is much more than the energy injection of the buoyancy. As $\Omega$ increases, the dominance of shear is reinforced. Additionally, the data of the buoyancy-dominated part in Regime III ($\Omega=6$) is not given in figure \ref{fig:Nu2}(b), because the accuracy of $\gamma$ is difficult to guarantee as $Nu_\omega$ is very close to $1$. The factor in front of the ratio of the Nusselt numbers, $31.79\Omega^2$, is large, therefore the buoyancy-driven flow ($|\gamma|<1$) is hard to influence $Nu_\omega$. From an alternative perspective, this can be attributed to the fact that the energy difference induced by angular velocity difference is much larger than the energy difference induced by density difference since the density variation is limited as $\alpha\Delta\ll 1$. 

\subsection{Flow structures}

To further investigate the dynamics behind the behaviors of heat and momentum transfer, we focus on the evolution of the flow structures. As the flow is stable in Regime II, the typical two-dimensional slice of temperature and angular velocity fields for Regimes I and III are visualized in figures \ref{fig:FlowStrucI} and \ref{fig:FlowStrucIII}, respectively. Next, we will explore and analyze the flow structures of the systems in Regimes I and III sequentially, to figure out how heat transfer is inhibited and facilitated, and how the system is stabilized.

\subsubsection*{Regime I}

\begin{figure}
	\centering
	\includegraphics[width=1.0\linewidth]{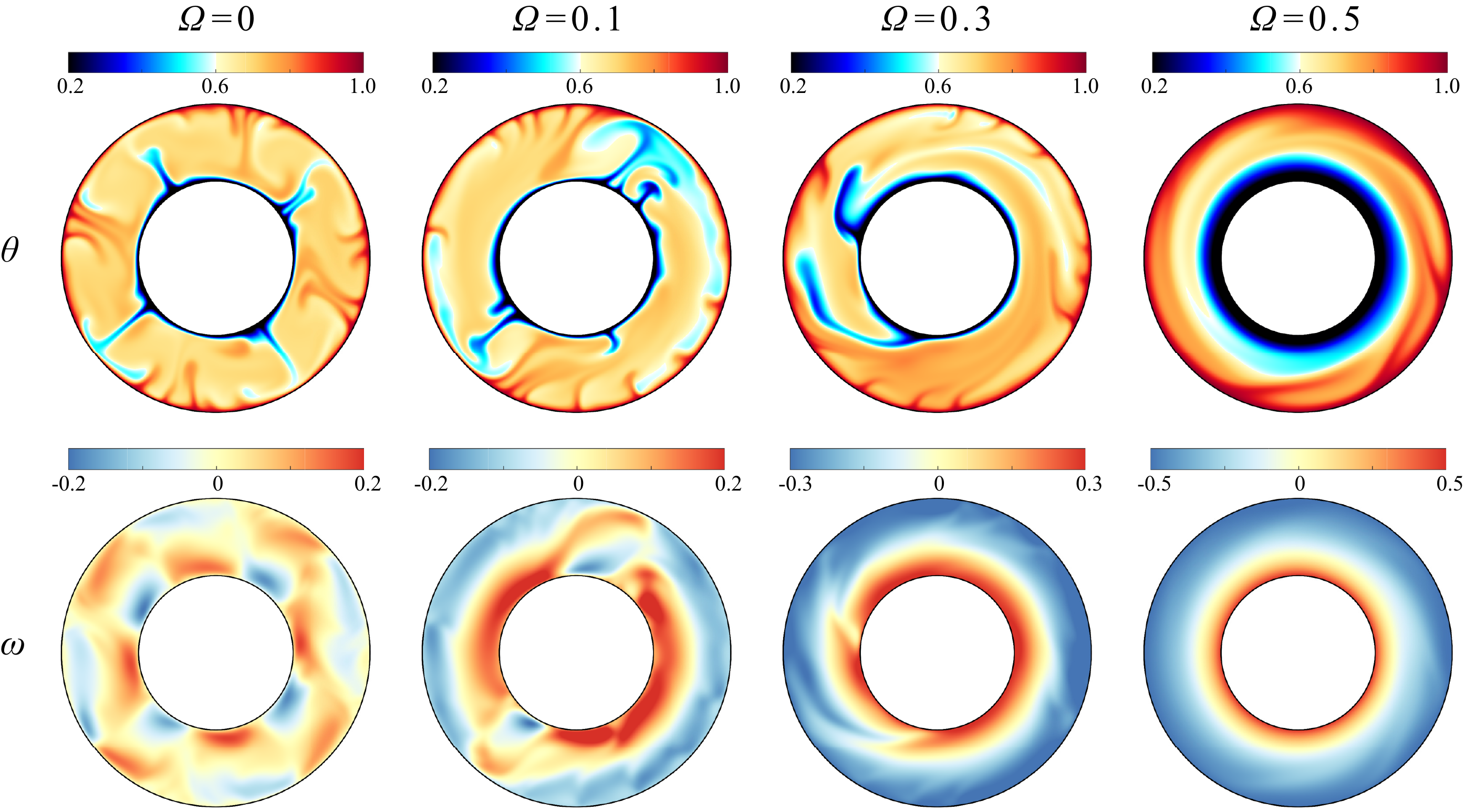}
        \captionsetup{justification=raggedright}
	\caption {Typical snapshots of the instantaneous temperature $\theta$ (the first row) and angular velocity $\omega=u_\varphi/r$ (the second row), at $\Omega=0, 0.1, 0.3, 0.5$ in Regime I. $Ra=10^6$.}
	\label{fig:FlowStrucI}
\end{figure}

In Regime I, the flow is quasi-two-dimensional under a large inversed Rossby number, $Ro^{-1}=20$. Therefore,  two-dimensional slices on the $r\varphi$ plane are taken for the temperature and angular velocity fields under different $\Omega$ at $Ra=10^6$, and the visualizations are displayed in figure \ref{fig:FlowStrucI}. Without shear, four pairs of convection rolls are formed, which can be clearly observed in the temperature and angular velocity snapshots. Due to the effect of Coriolis force, hot plumes detached from the outer cylinder turn right when rising, therefore one roll pair consists of a bigger roll on the right of the hot plume and a smaller roll on the left. This effect also results in the zonal flow in annular centrifugal RB convection, which is explained well by Jiang {\it et al.} \citep{jiang_supergravitational_2020} and Wang {\it et al.} \citep{wang_effects_2022}. The flow structure of the system is stable and the heat is continuously transferred through the plume with the LSC. Meanwhile, due to the asymmetry of the two rolls, the bigger roll takes more area and has a larger impact on the boundaries than the other roll, which pushes the inner cylinder to rotate anticlockwise and the outer cylinder to rotate clockwise. Hence, once the boundary is rotating slowly, a negative $Nu_\omega$ is generated.

It is conceivable that the flow structure shifts continuously with shear enhancement. As $\Omega$ increases to $0.1$, there is a notable alteration in the flow structure, whereby the quantity of convection roll pairs exhibits a decline. Only two pairs can be observed, and each pair consists of a large roll covering nearly half of the annulus and a quite small roll. These can be seen especially clearly in the angular velocity field. It can be speculated that the shear of two boundaries stretches the plumes and promotes the azimuthal motion of the plume in the shear direction. The stretched plumes under shear are also observed in sheared RB convection \citep{goluskin_convectively_2014,blass_flow_2020}. The flow motion induced by shear is in the same direction as the motion induced by the Coriolis force, including that hot plumes turn anticlockwise near the inner cylinder and the cold plumes turn clockwise near the outer cylinder; hence, the imposed shear increases the size of the large roll further. Meanwhile, the other roll becomes smaller. The number of roll pairs decreases as the system is not large enough to accommodate four roll pairs. Moreover, at this stage, the angular velocity of the fluid is still larger than the rotational speed of the boundary, and stronger than the angular velocity of the fluid without shear as well. This means, on the one hand, the azimuthal motion of the fluid is strengthened by shear; on the other hand, the shear is not strong enough and the azimuthal movement of the fluid driven by heat is still maintained. In the small rolls, the flow against the direction of boundary motion remains strong. Due to that the basic convection structures, the LSC and plumes, are maintained, and the heat transfer is reduced by only a small fraction, about $6\%$ compared to the heat transfer at $\Omega=0$.

As $\Omega$ continues to increase to $0.3$, one big roll covering near $3\pi/2$ is generated. In this case, the number of convection roll pairs is hard to define, as the roll against the direction of boundary motion nearly disappears. Actually, though we can see two strong cold plumes in the temperature field, there is no fluid rotating anticlockwise near the outer cylinder; while near the inner cylinder, only two small pieces of fluid rotate clockwise, against the direction of boundary motion. For all the fluid in the system, $|\omega|\le\Omega$, fluid can not rotate faster than the boundary wall and the $Nu_\omega$ turns positive but is still smaller than $1$. The rotating velocity of the boundaries exceeds the azimuthal velocity of the fluid induced by LSC. At this stage, the convection roll pairs are not maintained, while plumes remain working, transporting heat from one side to the other. Therefore, the heat transfer drops not too much but noticeably, reducing about $28\%$ compared to the heat transfer at $\Omega=0$.

As $\Omega$ continues to increase to $0.5$, near the marginal state, the flow is close to the laminar and nonvortical flow, and the plumes disappear in the end. Some temperature fluctuations are observed in the temperature field, while the angular velocity fluctuations are hard to recognize. At this stage, the convection and heat transfer are highly suppressed. Plumes, as important carriers of heat transfer, are hard to exist. The Nusselt number $Nu_h$ drops significantly, about $83\%$ compared to the $Nu_h$ at $\Omega=0$.

\begin{figure}
	\centering
	\includegraphics[width=0.9\linewidth]{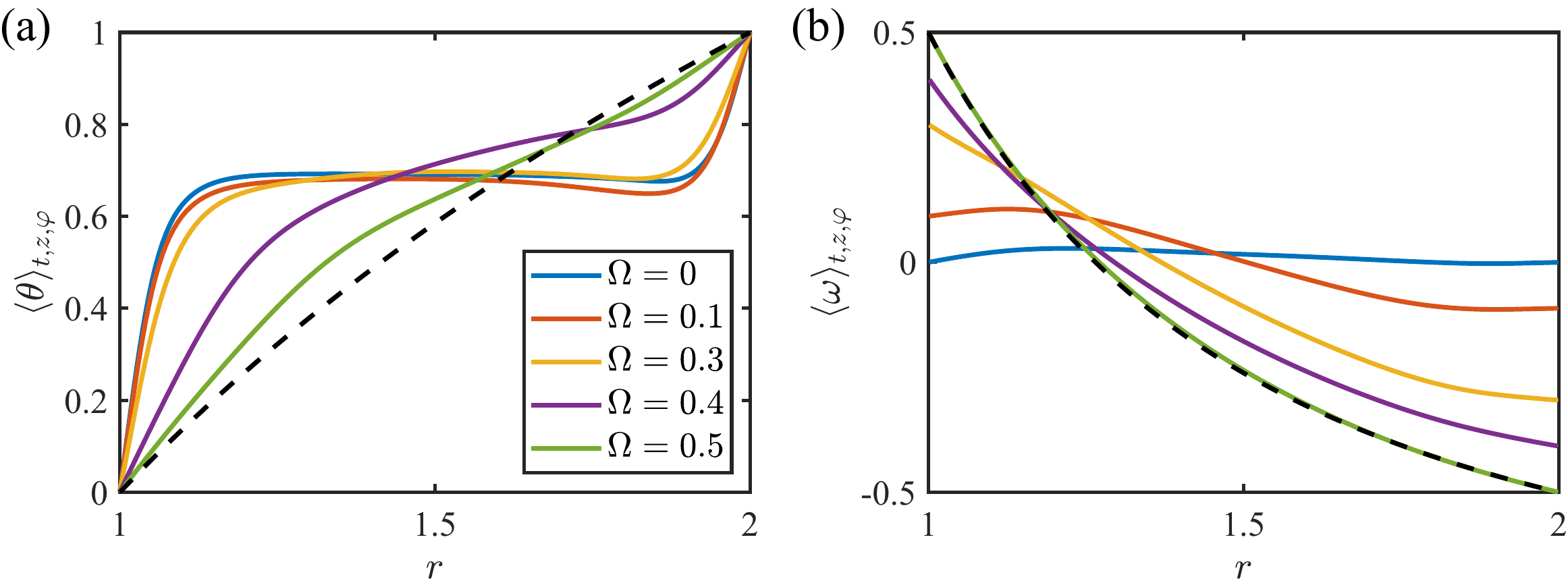}
        \captionsetup{justification=raggedright}
	\caption {Radial distribution of azimuthally, axially and time-averaged (\textit{a}) temperature $\langle\theta\rangle_{t,\varphi,z}$ and (\textit{b}) angular velocity $\langle\omega\rangle_{t,\varphi,z}$ profiles for different $\Omega$ in Regime I. $Ra=10^6$. The dashed lines mean the profiles of the laminar and nonvortical flow.}
	\label{fig:MeanflowI}
\end{figure}

The averaged temperature and angular velocity profiles under different $\Omega$ are shown in figure \ref{fig:MeanflowI}. A uniform bulk temperature $\theta_c$ is observed for $\Omega\le0.3$, which deviates from the arithmetic mean temperature of two boundaries $\theta_m=0.5$. The asymmetry temperature profiles without shear can be explained by the radially dependent gravity effects and the bulk temperature $\theta_c$ depends on the radius ratio $\eta$, which are discussed by Wang {\it et al.} \citep{wang_effects_2022}. The bulk temperature remains unchanged until $\Omega$ exceeds $0.3$. As the shear increases, the uniform bulk temperature vanishes and the temperature profile gradually evolves toward a laminar and nonvortical flow profile (the dashed line). For the angular velocity profile, a stage similar to the bulk temperature is not found. At $\Omega=0$, the averaged angular velocity is small as the asymmetry of the flow structure is not strong; as the shear is imposed, the averaged angular velocity varies from $\Omega$ at the inner cylinder to $-\Omega$ at the outer cylinder. It should be noted that when the shear is not strong (such as $\Omega=0.1$), the two extremums of the profile occur inside the cell rather than on the boundaries. While for $\Omega\ge0.3$, the extremums of the profile move to boundaries. As $\Omega$ increases, the angular velocity profile gradually evolves towards a laminar flow profile as well. At $\Omega=0.5$, it coincides almost exactly with the laminar and nonvortical flow profile, while the temperature profile still differs. Thus, we can see that as shear is enhanced, the changes in the angular velocity profile are not synchronized with the changes in the temperature profile; the former tends to occur earlier, either for the transition at $\Omega=0.3$ or at $\Omega=0.5$.

\subsubsection*{Regime III}

\begin{figure}
	\centering
	\includegraphics[width=0.6\linewidth]{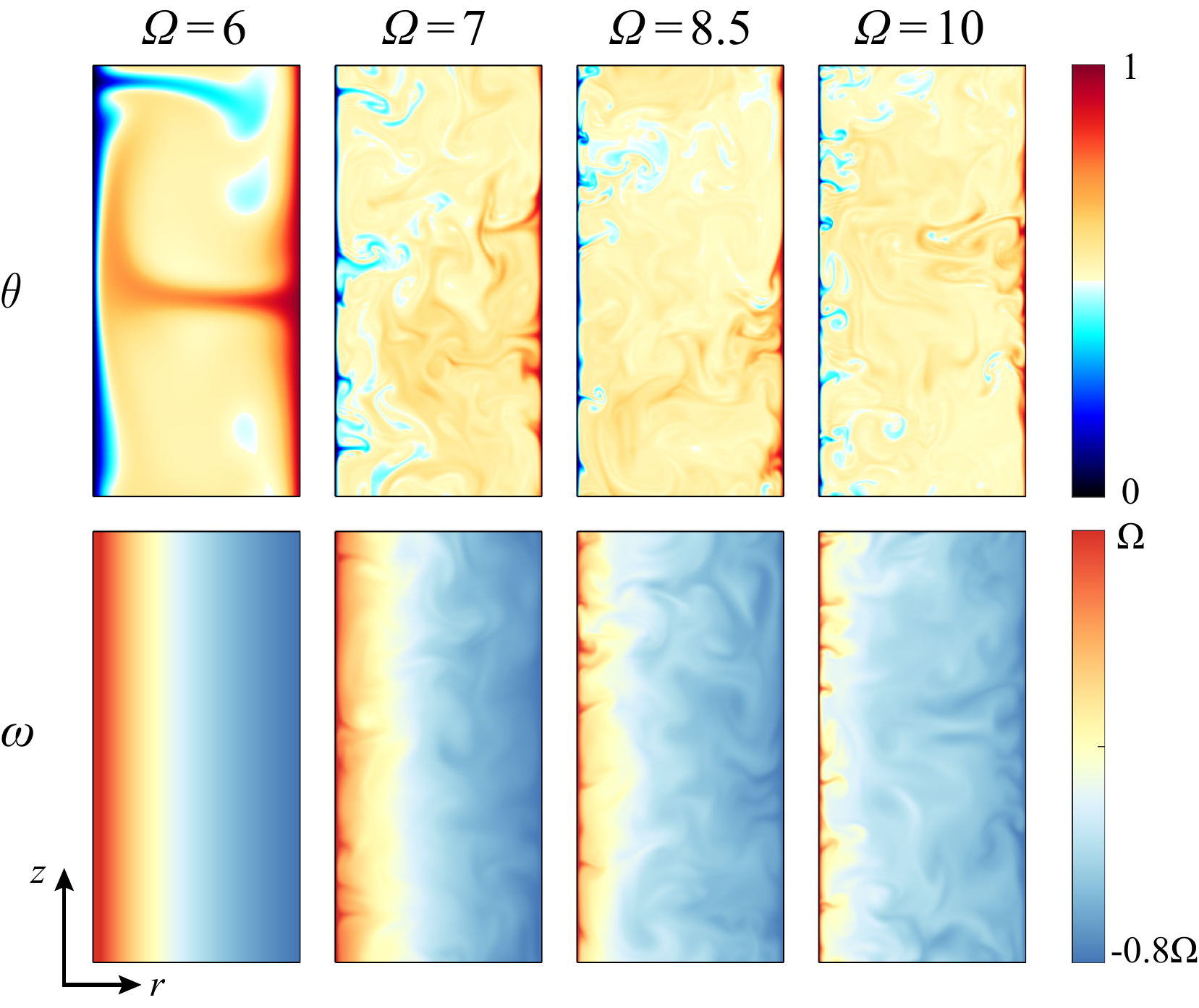}
        \captionsetup{justification=raggedright}
	\caption {Typical snapshots of the instantaneous temperature $\theta$ (the first row) and angular velocity $\omega=u_\varphi/r$ (the second row), at $\Omega=6, 7, 8.5, 10$ in Regime III. $Ra=10^6$. The corresponding Taylor numbers are $Ta=9.54\times 10^7, 1.30\times10^8, 1.91\times10^8, 2.65\times10^8$, respectively.}
	\label{fig:FlowStrucIII}
\end{figure}

In Regime III, though the flow is three-dimensional, the main structure is on the $rz$ plane \citep{grossmann_highreynolds_2016}. Therefore, two-dimensional slices are taken for the temperature and angular velocity fields, illustrated in figure \ref{fig:FlowStrucIII}. At $\Omega=6$, of which the inviscid TC system without buoyancy is at the marginal state, the flow shows a classical RB temperature field with a typical LSC. However, the angular velocity distribution is still almost laminar and nonvortical though convection occurs. As explained in subsection \ref{heatsec}, the convection driven by heat is hard to transfer angular momentum at large $\Omega$. With $\Omega$ increasing, the flow becomes more turbulent, with the angular velocity field better mixed and the plumes get thinner, meaning the transport efficiency is enhanced. At this stage, because the temperature acts as a passive scalar, the temperature fields and the angular velocity fields are highly relevant, which can be roughly seen by the highly similar plume morphology.

\begin{figure}
	\centering
	\includegraphics[width=0.9\linewidth]{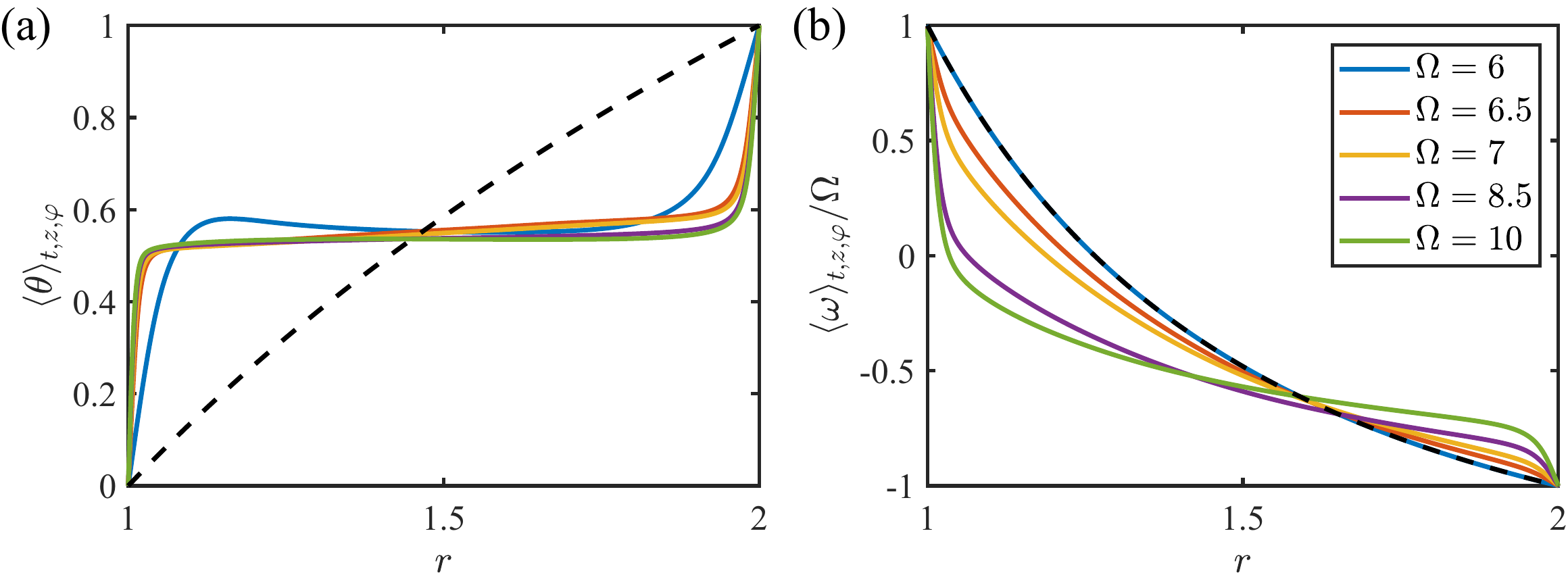}
        \captionsetup{justification=raggedright}
	\caption {Radial distribution of azimuthally, axially and time-averaged (\textit{a}) temperature $\langle\theta\rangle_{t,\varphi,z}$ and (\textit{b}) normalized angular velocity $\langle\omega\rangle_{t,\varphi,z}/\Omega$ profiles for different $\Omega$ in Regime III. $Ra=10^6$. The dashed lines mean the profiles of the laminar and nonvortical flow.}
	\label{fig:MeanflowIII}
\end{figure}

The averaged temperature and angular velocity profiles of Regime III are shown in figure \ref{fig:MeanflowIII}. A uniform bulk temperature is observed for all the cases, both the buoyancy-dominant case ($\Omega=6$) and the shear-dominated cases. Interestingly, this temperature does not change with $\Omega$, but is different from the arithmetic mean temperature $\theta_m=0.5$ and the bulk temperature in Regime I. As $\Omega$ increases, the profile becomes sharper at two boundaries, which also implies an improved efficiency of heat transfer. The angular velocity profile is in line with the laminar and nonvortical profile at $\Omega=6$, and then transits to a turbulent profile as the Taylor number is increasing. At $\Omega=10$, the outer cylinder is at rest, and the Taylor number $Ta=2.65\times10^8$, therefore the system is still in the classical regime of the TC flow \citep{ostilla-monico_exploring_2014,grossmann_highreynolds_2016}; the shear-dominated flow in the Regime III behaves similar to the TC flow.

\section{Conclusion}\label{sec4}

In the present study, we investigate the evolution of the flow regimes in an annular centrifugal Rayleigh-Bénard system with increasing imposed shear, which is represented by the angular velocity difference between the inner and outer cylinders, $\Omega$. At first, $Ri\ll 1$ under weak shear, the flow is buoyancy-dominated and manifests itself as quasi-two-dimensional convection on the $r\varphi$ plane. As $\Omega$ increases, the convection and heat transfer are depressed, and the flow finally turns to stable laminar and nonvortical flow. Through linear instability analysis, we notice that the marginal state depends on both $Ra$ and $\Omega$, formed by the energy balance among viscous dissipation, the work of buoyancy, and the work of inertial terms. At low $Ra\sim 10^3$, the marginal state mainly depends on $Ra$ and only a weak shear is required to stabilize the flow. At high $Ra\ge10^6$, the marginal state seems insensitive to $Ra$, locating at $Ri=\Omega^{-2}\sim O(1)$. In the parameter region $Ra\in[10^6,10^8]$, as shear increases, the growth rate of the instability is suppressed and the low azimuthal frequency modes become the main mode. Similarly, the heat transfer efficiency $Nu_h$ is also depressed gradually, and the number of roll pairs decreases as the shear enhances the asymmetry between the two opposing rolls and enlarges the bigger one. The plumes are stretched by the shear, and the temperature perturbations are also stretched in LSA as well. The LSA and DNS give many comparable findings. 

With the continued enhancement of shear, the flow becomes unstable again, but this time the flow is completely different from before. Initially, when the system has not crossed the marginal state derivated by Rayleigh's inviscid theory for pure TC flow, the instability of the flow is driven by buoyancy, and the thermal convection occurs on the $rz$ plane, similar to the classical RB convection. As $\Omega$ increases and the centrifugal instability takes over, the flow quickly becomes shear-dominated, with the temperature acting as a passive scalar. At this moment, the flow is similar to the TC flow, and still in the classical regime for our parameter sets. 

On the whole, as the flow transits from an annular centrifugal RB convection to a radially heated TC flow of only the inner cylinder rotating, it passes through three regimes. Regime I is mainly buoyancy dominated, Regime II is the stable regime while Regime III is mainly shear dominated. There is an obvious dividing line between the three regimes, that is the marginal state of the instability. Furthermore, the energy analysis through the exact relation implies more information. In Regime I, the energy injected by buoyancy is consumed by both turbulent dissipation and wall shear. As the shear increases, more percentages of injected energy are taken by the wall motion, and near the marginal state, it takes about $50\%$. Therefore, shear plays an important role in Regime I as well, diverting buoyant injections of energy and inhibiting the convection. Meanwhile, on the other end of the stable regime, buoyancy drives the flow at first. Convection is observed in the temperature profile while the angular velocity profile is still close to the profile in the laminar and nonvortical case. When near the marginal state, both shear and buoyancy are important, and the exact relation (the equation (\ref{NuEqu})) is a good approach to analyzing the flow characteristics. 

The traditional Rayleigh-Bénard system sheared by wall motion in a plane Couette flow is remarkably close to our system. Blass and Yerragolam have conducted a series of meaningful studies on this topic recently \citep{blass_flow_2020, blass_effect_2021, yerragolam_passive_2022, yerragolam_how_2022}. As shear increases from $0$, the flow regime moves from the buoyancy-dominated regime, through the transitional regime, and finally into the shear-dominated regime. A non-monotonic progression of $Nu$ is observed as well, with a minimum in the transitional regime, where $Ri\sim 1$. These observations are consistent with ours, suggesting that the dominance of buoyancy or shear is a key factor in determining the flow state. Moreover, the coupling effects of buoyancy and shear on the heat transfer are also close between the two systems. However, due to the effect of curvature and rotation, the evolution of flow structures in the sheared ACRBC exhibits a lot of differences. The appearance of the stable regime is a very special finding. Meanwhile, because of the differences between the Taylor-Couette flow and the plane Couette flow, the flow in the shear-dominated regime is quite dissimilar. The similarities and differences between the sheared ACRBC and sheared RB are interesting points that deserve our future research.

The current system properly couples the TC flow with the RB convection through a couple of differential rotation centrifugal cylinders and is a feasible approach to analyze the coupling effect of shear and buoyancy. Limited by the parameters space, there are plenty of questions to be further explored. In the region of high Rayleigh number, how does shear affect the $Nu\sim Ra^\gamma$ scaling? Meanwhile, the effect of other parameters on the stable region, including $Ro^{-1}$ and $Pr$, is also an interesting project. Moreover, some studies show that as the radius ratio $\eta$ increases and approaches $1$, the convection in the annulus system behaves more and more similar to the convection between two planes \citep{pitz_onset_2017,wang_effects_2022}. As we know, there is no stable regime for a sheared RB convection between two horizontal planes \citep{blass_flow_2020}. The disappearance of the stable regime is an attractive phenomenon that is deserved to be explored in the future.\clearpage 

\backsection [Acknowledgements] {We thank Detlef Lohse and Hechuan Jiang for the helpful discussions.}

\backsection [Funding]{This work was supported by the National Natural Science Foundation of China under grant no. 11988102, and the New Cornerstone Science Foundation through the XPLORER PRIZE}.

\backsection[Declaration of interests] {The authors report no conflict of interest.}

\appendix

\section{Simulation parameters}\label{Tab1}

\begin{table}
  \begin{center}
\def~{\hphantom{0}}
    \begin{tabular}{cccccccccccc}
	No.&$Ra$ &$\Omega$ &$Ta$ & $N_\varphi\times N_z\times N_r$& $\Delta_g/\eta_K$ & $Nu_h$& $\epsilon_{Nu_h}$ &$Nu_\omega$ & $\epsilon_{Nu_\omega}$ & $\Gamma$ & $\phi_0$ \\
	$1   $&$ 1.0\times 10^6 $&$ 0        $&$ 0 $&$ 1024\times 32\times 128 $&$ 0.29            $&$ 7.30   $&$ 0.30\%  $&$ -                       $&$ -                        $&$ 0.25  $&$ 1    $ \\
	$2   $&$ 1.0\times 10^6 $&$ 0.01     $&$ 2.65\times 10^2 $&$ 1024\times 32\times 128 $&$ 0.29            $&$ 6.96   $&$ 0.86\%  $&$ -7.22                     $&$ 2.83\%                     $&$ 0.25  $&$ 1    $ \\
	$3   $&$ 1.0\times 10^6 $&$ 0.1      $&$ 2.65\times 10^4 $&$ 1024\times 32\times 128 $&$ 0.28            $&$ 6.85   $&$ 0.99\%  $&$ -0.39                     $&$ 2.29\%                     $&$ 0.25  $&$ 1    $ \\
	$4   $&$ 1.0\times 10^6 $&$ 0.2      $&$ 1.06\times 10^5 $&$ 1024\times 32\times 128 $&$ 0.27            $&$ 6.35   $&$ 0.49\%  $&$ 0.18                      $&$ 3.84\%                     $&$ 0.25  $&$ 1    $ \\
	$5   $&$ 1.0\times 10^6 $&$ 0.3      $&$ 2.38\times 10^5 $&$ 1024\times 32\times 128 $&$ 0.24            $&$ 5.27   $&$ 0.22\%  $&$ 0.54                      $&$ 2.73\%                     $&$ 0.25  $&$ 1    $ \\
	$6   $&$ 1.0\times 10^6 $&$ 0.35     $&$ 3.25\times 10^5 $&$ 1024\times 32\times 128 $&$ 0.20            $&$ 3.08   $&$ 2.79\%  $&$ 0.78                      $&$ 3.44\%                     $&$ 0.25  $&$ 1    $ \\
	$7   $&$ 1.0\times 10^6 $&$ 0.4      $&$ 4.24\times 10^5 $&$ 1024\times 32\times 128 $&$ 0.17            $&$ 2.09   $&$ 3.24\%  $&$ 0.91                      $&$ 3.77\%                     $&$ 0.25  $&$ 1    $ \\
	$8   $&$ 1.0\times 10^6 $&$ 0.5      $&$ 6.62\times 10^5 $&$ 1024\times 32\times 128 $&$ 0.12            $&$ 1.28   $&$ 2.02\%  $&$ 0.98                      $&$ 8.62\%                     $&$ 0.25  $&$ 1    $ \\
	$9   $&$ 1.0\times 10^7 $&$ 0        $&$ 0 $&$ 1024\times 32\times 128 $&$ 0.62            $&$ 13.29  $&$ 0.36\%  $&$ -                       $&$ -                        $&$ 0.25  $&$ 1    $ \\
	$10  $&$ 1.0\times 10^7 $&$ 0     $&$ 0 $&$ 1536\times 32\times 192 $&$ 0.41            $&$   13.28	$&$   0.54\%      $&$    -                   $& $      -           $&$   0.25    $&$   1   $ \\
	$11  $&$ 1.0\times 10^7 $&$ 0.01     $&$ 2.65\times 10^3 $&$ 1024\times 32\times 128 $&$ 0.62            $&$ 12.81 	$&$ 0.10\%  $&$ -18.49         $&$ 1.87\%                     $&$ 0.25  $&$ 1    $ \\
	$12  $&$ 1.0\times 10^7 $&$ 0.01     $&$ 2.65\times 10^3 $&$ 1536\times 32\times 192 $&$ 0.41            $&$ 12.73 	$&$ 0.30\%  $&$ -18.42         $&$ 0.38\%                     $&$ 0.25  $&$ 1    $ \\
	$13  $&$ 1.0\times 10^7 $&$ 0.1      $&$ 2.65\times 10^5 $&$ 1024\times 32\times 128 $&$ 0.59            $&$ 12.42  $&$ 0.57\%  $&$ -1.55                     $&$ 0.03\%                     $&$ 0.25  $&$ 1    $ \\
	$14  $&$ 1.0\times 10^7 $&$ 0.2      $&$ 1.06\times 10^6 $&$ 1024\times 32\times 128 $&$ 0.58            $&$ 12.20  $&$ 1.55\%  $&$ -0.47                     $&$ 2.42\%                     $&$ 0.25  $&$ 1    $ \\
	$15  $&$ 1.0\times 10^7 $&$ 0.3      $&$ 2.38\times 10^6 $&$ 1024\times 32\times 128 $&$ 0.54            $&$ 10.33  $&$ 1.01\%  $&$ 0.35                      $&$ 1.37\%                     $&$ 0.25  $&$ 1    $ \\
	$16  $&$ 1.0\times 10^7 $&$ 0.4      $&$ 4.24\times 10^6 $&$ 1024\times 32\times 128 $&$ 0.47            $&$ 7.51   $&$ 1.09\%  $&$ 0.50                      $&$ 3.15\%                     $&$ 0.25  $&$ 1    $ \\
	$17  $&$ 1.0\times 10^7 $&$ 0.45     $&$ 5.36\times 10^6 $&$ 1024\times 32\times 128 $&$ 0.34            $&$ 3.12   $&$ 0.74\%  $&$ 0.84                      $&$ 0.86\%                     $&$ 0.25  $&$ 1    $ \\
	$18  $&$ 1.0\times 10^8 $&$ 0        $&$ 0 $&$ 2048\times 32\times 256 $&$ 0.66            $&$ 26.06  $&$ 0.04\%  $&$ -                       $&$ -                        $&$ 0.125 $&$ 1    $ \\
	$19  $&$ 1.0\times 10^8 $&$ 0     $&$ 0 $&$ 2560\times 32\times 270 $&$     0.58          $&$   26.12    $&$    1.02\%     $& $     -                $&$       -          $&$   0.125    $&$   1   $ \\
	$20  $&$ 1.0\times 10^8 $&$ 0.01     $&$ 2.65\times 10^4 $&$ 2048\times 32\times 256 $&$ 0.65            $&$ 24.70  $&$ 0.73\%  $&$ -25.07                    $&$ 4.40\%                     $&$ 0.125 $&$ 1    $ \\
	$21  $&$ 1.0\times 10^8 $&$ 0.1      $&$ 2.65\times 10^6 $&$ 2048\times 32\times 256 $&$ 0.64            $&$ 25.11  $&$ 0.04\%  $&$ -1.33                     $&$ 5.18\%                     $&$ 0.125 $&$ 1    $ \\
	$22  $&$ 1.0\times 10^8 $&$ 0.2      $&$ 1.06\times 10^7 $&$ 2048\times 32\times 256 $&$ 0.60            $&$ 21.24  $&$ 1.69\%  $&$ -0.63                     $&$ 4.03\%                     $&$ 0.125 $&$ 1    $ \\
	$23  $&$ 1.0\times 10^8 $&$ 0.3      $&$ 2.38\times 10^7 $&$ 2048\times 32\times 256 $&$ 0.57            $&$ 17.75  $&$ 0.63\%  $&$ 0.14                      $&$ 5.09\%                     $&$ 0.125 $&$ 1    $ \\
	$24  $&$ 1.0\times 10^8 $&$ 0.4      $&$ 4.24\times 10^7 $&$ 2048\times 32\times 256 $&$ 0.57            $&$ 17.35  $&$ 1.69\%  $&$ 0.56                      $&$ 6.19\%                    $&$ 0.125 $&$ 1    $ \\
	$25  $&$ 1.0\times 10^8 $&$ 0.5      $&$ 6.62\times 10^7 $&$ 2048\times 32\times 256 $&$ 0.42            $&$ 10.11  $&$ 0.71\%  $&$ 0.40                      $&$ 4.47\%                     $&$ 0.125 $&$ 1    $ \\
	$26  $&$ 1.0\times 10^6 $&$ 6        $&$ 9.54\times 10^7 $&$ 384\times 256\times 256 $&$ 0.19            $&$ 6.50   $&$ 2.10\%  $&$ 1.00                      $&$ 0                          $&$ 2/3\pi $&$ 1/4 $ \\
	$27  $&$ 1.0\times 10^6 $&$ 6.5      $&$ 1.12\times 10^8 $&$ 384\times 256\times 256 $&$ 0.82            $&$ 19.97  $&$ 0.67\%  $&$ 2.42                      $&$ 1.41\%                     $&$ 2/3\pi $&$ 1/4 $ \\
	$28  $&$ 1.0\times 10^6 $&$ 7        $&$ 1.30\times 10^8 $&$ 384\times 256\times 256 $&$ 1.01            $&$ 23.58  $&$ 0.81\%  $&$ 3.90                      $&$ 0.79\%                     $&$ 2/3\pi $&$ 1/4 $ \\
	$29  $&$ 1.0\times 10^6 $&$ 7        $&$ 1.30\times 10^8 $&$ 768\times 256\times 256 $&$ 1.01            $&$ 23.38  $&$ 1.00\%  $&$ 3.88                      $&$ 1.35\%                     $&$ 2/3\pi $&$ 1/2  $ \\
	$30  $&$ 1.0\times 10^6 $&$ 8.5      $&$ 1.91\times 10^8 $&$ 432\times 324\times 324 $&$ 1.35            $&$ 28.61  $&$ 0.74\%  $&$ 7.27                      $&$ 0.77\%                     $&$ 2/3\pi $&$ 1/4 $ \\
	$31  $&$ 1.0\times 10^6 $&$ 10       $&$ 2.65\times 10^8 $&$ 432\times 324\times 324 $&$ 1.61            $&$ 31.63  $&$ 1.75\%  $&$ 9.98                      $&$ 1.25\%                     $&$ 2/3\pi $&$ 1/4 $ \\
	$32  $&$ 1.0\times 10^6 $&$ 10       $&$ 2.65\times 10^8 $&$ 432\times 324\times 432 $&$ 1.26 
	$&$   30.83	$&$     0.71\%    $& 	$  9.83        $&$        0.36\%           $&$ 2/3\pi $&$ 1/6 $\\
    \end{tabular}
    \captionsetup{justification=raggedright}
    \caption{ Simulation parameters. The columns dispaly the Rayleigh number $Ra$, the non-dimensional rotational speed difference $\Omega$, the Taylor number $Ta$, the resolution employed, the maximum grid spacing $\Delta_g$ compared with the Kolmogorov scale estimated by the global criterion $\eta_K=(\nu^3/\varepsilon)^{1/4}$, the calculated Nusselt numbers $Nu_h$, $Nu_\omega$ and their relative difference of two halves $\epsilon_{Nu}=|(\langle Nu\rangle_{0-T/2}-\langle Nu\rangle_{T/2-T})/(Nu-1)|$, the aspect ratio $\Gamma=H/L$, and the reduced azimuthal domain $\phi_0$. $\varepsilon$ is the mean energy dissipation rate calculated by the equation (\ref{NuEqu}) of $C\approx 1$. Note that the azimuthal resolution corresponds to the resolution of the segment. For example, $N_\varphi=768$ for $\phi_0=1/2$ means $768$ points for one half of the annulus.}
\label{tab:my-table}
  \end{center}
\end{table}



\end{document}